\documentclass[aps, prl, reprint, superscriptaddress, preprintnumbers]{revtex4-2}

\usepackage{float}
\usepackage{physics}
\usepackage{amssymb}

\usepackage{hyperref}
\usepackage[capitalise]{cleveref}

\newcommand{\dbar}{%
  \mathchar'26\mkern-12mu \text{d}%
}

\NewDocumentCommand{\ddbar}{o m}{%
  \IfNoValueTF{#1}
    {\dbar{#2}\,}
    {\dbar^{#1}{#2}\,}
}

\usepackage{graphicx}

\begin{document}

\preprint{\vbox{
\hbox{JLAB-THY-25-4360}
}}

\title{Real-time Estimators for Scattering Observables:\\
\small A full account of finite-volume errors for quantum simulation}

\author{Ivan M. Burbano}
\email{ivan\_burbano@berkeley.edu}
\affiliation{Department of Physics, University of California, Berkeley, CA 94720, USA}
\affiliation{Nuclear Science Division, Lawrence Berkeley National Laboratory, Berkeley, CA 94720, USA}
\affiliation{Physics Division, Lawrence Berkeley National Laboratory, Berkeley, CA 94720, USA}
\author{Marco A. Carrillo}
\email{mcarr020@odu.edu}
\affiliation{Department of Physics, Old Dominion University, Norfolk, Virginia 23529, USA}
\affiliation{Thomas Jefferson National Accelerator Facility, 12000 Jefferson Avenue, Newport News, Virginia 23606, USA}
\author{Rana Urek}
\email{urek@berkeley.edu}
\affiliation{Department of Physics, University of California, Berkeley, CA 94720, USA}
\affiliation{Nuclear Science Division, Lawrence Berkeley National Laboratory, Berkeley, CA 94720, USA}
\author{Anthony N. Ciavarella}
\email{anciavarella@lbl.gov}
\affiliation{Physics Division, Lawrence Berkeley National Laboratory, Berkeley, CA 94720, USA}
\author{Ra\'ul A. Brice\~no}
\email{rbriceno@berkeley.edu}
\affiliation{Department of Physics, University of California, Berkeley, CA 94720, USA}
\affiliation{Nuclear Science Division, Lawrence Berkeley National Laboratory, Berkeley, CA 94720, USA}

\begin{abstract}

The real-time correlators of quantum field theories can be directly probed through new approaches to simulation, such as quantum computing and tensor networks.
This provides a new framework for computing scattering observables in lattice formulations of strongly interacting theories, such as lattice quantum chromodynamics.
In this paper, we prove that the proposal of real-time estimators of scattering observables is universally applicable to all scattering observables of gapped quantum field theories.
All finite-volume errors are exponentially suppressed, and the rate of this suppression is controlled by the regulator considered, namely, a displacement of the spectrum of the theory into the complex plane. 
A partial restoration of Lorentz symmetry by averaging over different boosts gives an additional suppression of finite volume errors. 
Our results also apply to the simulation of wavepacket scattering, where a similar averaging is performed to construct the wavepackets that regulate the finite volume effects. This result represents a necessary key step towards determining a broad class of scattering observables via quantum computing that are currently inaccessible via classical computing. Such observables are relevant for various applications, including hadron spectroscopy, hadron structure, and precision tests of the Standard Model.
We also comment on potential applications of our results to traditional computational schemes.

\end{abstract}

\maketitle

\textbf{\textit{Introduction.}}---The discovery potential of nuclear and particle experiments around the world hinges on precise and accurate predictions from the Standard Model of particle physics. Two flagship experiments exemplify this need: the Deep Underground Neutrino Experiment (DUNE)~\cite{DUNE:2016hlj, DUNE:2020fgq}, aimed at providing further constraints on beyond-the-standard-model observables, and the future Electron-Ion Collider (EIC) experiment~\cite{Accardi:2012qut,AbdulKhalek:2022hcn}, which seeks to map the 3D structure of protons and light nuclei, including gluon distributions and quark-gluon correlations. Whether the desire is to remove the standard model ``background," as is the case in DUNE, or to obtain confirmation of hadronic observables, as in the EIC, present and future scattering experiments require a robust theoretical program aimed at determining the direct consequences of the contributions of quantum chromodynamics (QCD) to scattering observables. 

Despite QCD having been identified as the fundamental theory for the strong nuclear force for over 50 years, its non-perturbative nature has greatly restricted the class of observables that can be directly obtained from the theory. Lattice QCD calculations performed on classical computers have helped lift this by providing direct and indirect determinations of low-energy observables, including scattering amplitudes~\cite{Luscher:1985dn, Luscher:1986pf, Luscher:1990ux}. The present paradigm for extracting scattering observables via lattice QCD requires rather sophisticated non-perturbative formalisms to map the observables that are directly accessible via lattice QCD, namely finite-Euclidean spacetime correlators, to desired scattering quantities (see Refs.~\cite{Briceno:2017max, Hansen:2019nir,Mai:2021lwb, Bulava:2022ovd} for reviews on this topic). The complexity of the derivation and implementation of this formalism quickly grows with the energy of the reactions. The cutting edge of these studies involves kinematics where at most three particles can go on their mass shell~\footnote{which happens when the energy of the system is equal to or greater than the sum of the masses of the three particles}(see Refs.~\cite{Dawid:2025zxc,  Jackura:2022gib, Hansen:2020otl, Briceno:2024ehy,Mai:2018djl,Mai:2017bge} for recent formal and numerical developments in making this possible). 
In fact, it is known that scattering is a BQP-complete problem~\cite{Jordan:2017lea}, meaning that the ability to compute scattering amplitudes efficiently is equivalent to being able to simulate a generic quantum computation efficiently. Therefore, it is likely that an efficient framework that circumvents this bottleneck will require quantum computers. 

With this in mind, we review the minimal steps needed to realistically access scattering observables of strongly interacting quantum field theories via quantum computers. Given that the primary advantage that quantum hardware may present over classical computing is the access to real-time observables, we focus our attention on these quantities. Of course, the whole program hinges on the availability of quantum hardware. The remaining ingredients include,
\begin{enumerate}
    \item a systematically improvable definition of scattering estimators~\cite{Briceno:2020rar, Briceno:2021aiw, Carrillo:2024chu},
    \item a digitization of the theory onto the quantum device~\cite{Raychowdhury:2019iki, Kadam:2022ipf, Kadam:2024ifg, Ciavarella:2021nmj, DAndrea:2023qnr, Grabowska:2024emw, Burbano:2024uvn, Bergner:2024qjl, Ciavarella:2024fzw, Ciavarella:2025bsg, Assi:2024pdn}, 
    \item an efficient circuit for the evaluation of time-dependent observables~\cite{Charles:2023zbl, Bennewitz:2025nhz, Davoudi:2025rdv, Schuhmacher:2025ehh, Farrell:2025nkx, Ingoldby:2025bdb, Yusf:2024igb, Chai:2025qhf, Abel:2025zxb, Sharma:2023bqu, Wu:2024adk},
    \item an error correcting algorithm, if needed~\cite{Rajput:2021trn, Yao:2025cxs, Halimeh:2021vzf, Halimeh:2021lnv, Chen:2022dox, Stryker:2018efp, Spagnoli:2024mib, Carena:2024dzu, Faist:2019ahr}.
\end{enumerate}
The majority of the field working towards this goal has focused on the last three items in this list (for recent reviews on these topics see Refs.~\cite{Halimeh:2025vvp, Davoudi:2025kxb, Bauer:2025nzf, DiMeglio:2023nsa, Bauer:2022hpo, Davoudi:2022bnl, Catterall:2022wjq}).
In this work, we focus on the first item instead.
In brief, we prove that computational frameworks intended to study scattering through real-time correlators are systematically improvable. We further display the asymptotic dependence of the error in these computations on the system size. This proof is independent of the details in items 2-4 in the list above. 
Moreover, while items 2-4 are dependent of the particular theory one has in mind, the discussion in this paper, while highly motivated by QCD, is completely general and applies uniformly to the study of any scattering observable in an arbitrary quantum field theory.

\begin{figure*}
    \centering
    \includegraphics[height = 0.27\textheight]{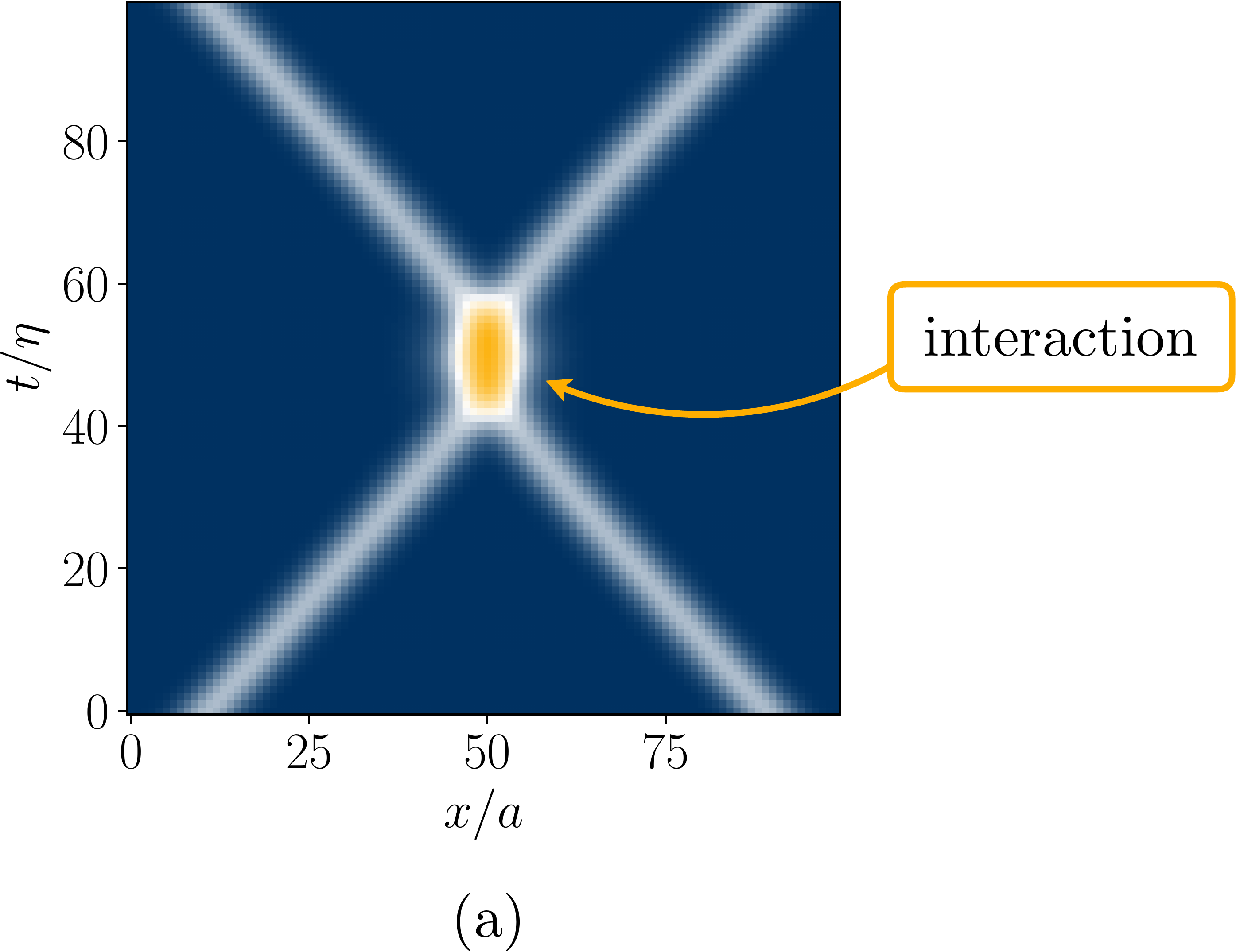}
    \hspace{0.01\textwidth}
    \includegraphics[height = 0.27\textheight]{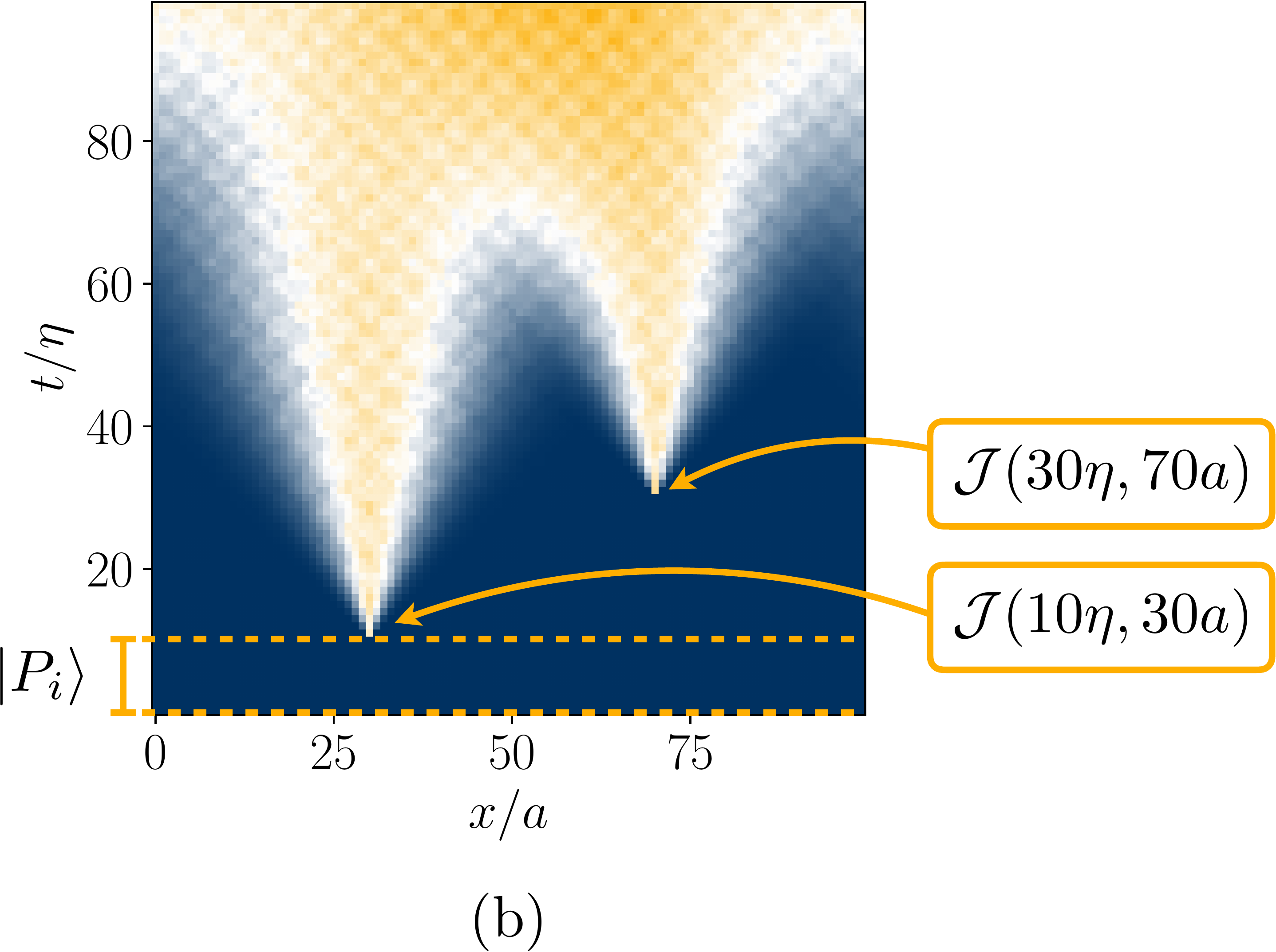}
    \caption{
    Graphical description of the wavepacket (a) and RESOs (b) approaches for two-particle elastic scattering. 
    In these, $a$ and $\eta$ are lattice spacings in the spatial and temporal directions.
    (a) Initially, two non-overlapping one-particle states are created. 
    Then, through time evolution, the states overlap and the scattering process occurs. 
    Finally, through further evolution, two non-overlapping wavepackets are recovered. 
    (b) Initially, a definite-momentum one-particle state $\ket{P_i}$ is prepared. 
    Then, two separated currents are applied to the state. 
    Two-particle observables are recovered through the overlap of the resulting state with another definite-momentum one-particle state.
    }
    \label{fig:wavefunction_v_reso}
\end{figure*}

To precisely state our results, let us describe some of the current approaches to study lattice QCD with the help of quantum computers \cite{Jordan:2012xnu,Jordan:2011ci,Jordan:2014tma, Jha:2024jan,Gustafson:2019mpk,Gustafson:2021imb,Parks:2022kdb, Ciavarella:2020vqm, Briceno:2020rar, Briceno:2021aiw, Briceno:2023xcm, Iadecola:2023uti, Carrillo:2024chu, Chai:2023qpq, Chai:2025qhf}. 
We first consider the seminal work of Jordan, Lee, and Preskill~\cite{Jordan:2012xnu,Jordan:2014tma}, where a protocol to simulate scattering in a scalar field theory using wavepackets was proposed.
This procedure, which is qualitatively depicted in \cref{fig:wavefunction_v_reso}(a), requires the construction of two or more wavepackets created at asymptotic distances in the far past, which are time evolved to interact with one another. From the time profile of the overlaps of these wavefunctions, one can determine, for example, their scattering phase shift. Since that work, variational techniques have been developed to prepare vacuum states and particle wavepackets in gauge theories on quantum computers at scale~\cite{Farrell:2023fgd,Farrell:2024fit,Ciavarella:2024lsp,Farrell:2024mgu,Gustafson:2024bww,Davoudi:2024wyv}.
These techniques have also been extended to simulate the dynamics of scattering in field theories variationally~\cite{Liu:2021otn,Zemlevskiy:2024vxt,Farrell:2025nkx}.
Recently, scattering of hadrons in $\mathbb{Z}_2$ and $\mathrm{U}(1)$ gauge theories with $D=1+1$ has been explored using these methods~\cite{Davoudi:2025rdv,Schuhmacher:2025ehh}.
In all of these calculations, the variational parameters describing the states have an exponentially fast convergence in system size.
As these parameters describe every aspect of the states, these exploratory studies suggest that the finite volume effects in wavepacket scattering are exponentially small.
Although there is still no formal proof of this, the formalism presented here suggests this has to be the case.

An alternative method, proposed in~\cite{Briceno:2020rar, Briceno:2021aiw, Briceno:2023xcm, Carrillo:2024chu}, provides a universal approach for obtaining any scattering amplitude.
Its input are time-ordered correlators that could be evaluated from quantum computers.
We will call this proposal the Real-time Estimators for Scattering Observables (RESOs) approach.
The starting point in this formulation is the construction of a scattering amplitude $\mathcal{T}$ in terms of real-time correlators as  
\begin{equation} \label{eq:TM_gen} 
    \begin{aligned}
        \mathcal{T} &=\lim_{\varepsilon\to0}\lim_{\mathcal{V}\to\infty} \int_{\mathcal{V}} \prod_{n} \qty(\dd[D]{x_n} e^{iq_n\cdot x_n-\varepsilon |t_n| })\times \\
        &\hphantom{\int_{\mathcal{V}} \prod_{n} \qty(\dd[D]{x_n} )}\mel{ P_f }{{\rm T}\qty[\prod_{n} \mathcal{J}_n(x_{n})]}  { P_i }, 
\end{aligned}
\end{equation}
where the initial/final states $\ket{P_{i/f}}$ are either one-particle states or the vacuum in a $D=1+d$-dimensional spacetime volume $\mathcal{V}=T\times L^d$.
The operators $\mathcal{J}_n(x_n)$ are currents with momentum $q_n$ that can either serve as physical probes of a hadronic system or as interpolating operators to access multi-hadron scattering amplitudes using the Lehmann, Symanzik, and Zimmermann reduction formula. In \cref{fig:wavefunction_v_reso}(b), we give a qualitative depiction of one example of the matrix elements appearing in the integrand above. This example shows a plane wave state being struck by two local currents separated in time.

These time-ordered correlation functions can be computed using existing quantum algorithms~\cite{PhysRevLett.113.020505,wang2025computingntimecorrelationfunctions}.
This formulation crucially depends on two regularization ingredients:
\begin{enumerate}
    \item The introduction of a small but non-zero parameter $\varepsilon > 0$, which pushes the spectrum of the theory into the complex plane.
    \item The averaging over different external kinematics, also known as boost-averaging, which enhances the relativistic symmetry of the final result.
\end{enumerate}

Ultimately, we are interested in constraining scattering observables, which require the notion of asymptotic states. This poses an issue since any realistic quantum device will introduce an infrared (IR) cutoff, such as a finite volume with periodic boundary conditions. Given that asymptotic states can not be rigorously defined in a finite spatial volume, it is critical to understand the systematics of this constraint (for a concrete example of these effects in the context of quantum computation, see Ref. \cite{Farrell:2022vyh}). We will thus focus on specifying the effects of the two regulators mentioned above, $\varepsilon$ and boost averaging, on the finite-volume corrections of real-time correlators.
Schematically, we will show that our first regulator induces an exponential suppression which, at worst, is of the form $e^{-\varepsilon L}$ (a more precise statement is in \cref{eq:eps_suppression}).
While this is enough to show that, in principle, the RESOs approach converges universally across all scattering observables, in practice, this suppression is slow.
We thus further show that boost-averaging further suppresses the finite-volume errors by the characteristic function of the averaging measure.
This was seen through numerical explorations in~\cite{Briceno:2020rar,Carrillo:2024chu} for specific kinematic thresholds.
Our findings in this work prove that this suppression occurs in general. 

Before continuing with our main discussion, we note that our results are also useful in the wavepacket approach.
In that approach, the introduction of wavepackets is fundamental to ensure a sensible infinite volume limit.
The contribution from multiple momentum modes present in these results in a suppression of finite-volume effects similar to that due to boost-averaging.
This is elaborated on in the End Matter.

\textbf{\textit{Generic diagrams in a finite volume.}}---
To understand the finite-volume corrections for an arbitrary reaction in a generic quantum field theory~\footnote{Given our interest in finite-volume artifacts, we will not discuss ultraviolet (UV) divergences in this work. Since both finite-volume and infinite-volume diagrams share the same UV structure, finite-volume corrections are UV finite.}, it is sufficient to consider the IR degrees of freedom for that theory. 
For QCD, this would be an effective theory of hadrons.
In the effective theory, the time-ordered correlators used in this work will be expressed as a sum over Feynman diagrams.
To ensure that our result is general, we will consider arbitrary classes of diagrams, making no assumptions on the underlying dynamics of the theory.
However, to keep notation simple, in the main body of this text, we will write as if our theory were comprised of a single scalar field whose interactions only involve scalar couplings.
We explain in the End Matter how the conclusions are unaffected by these assumptions.

\begin{figure*}
    \centering
    \includegraphics[width=0.35\linewidth]{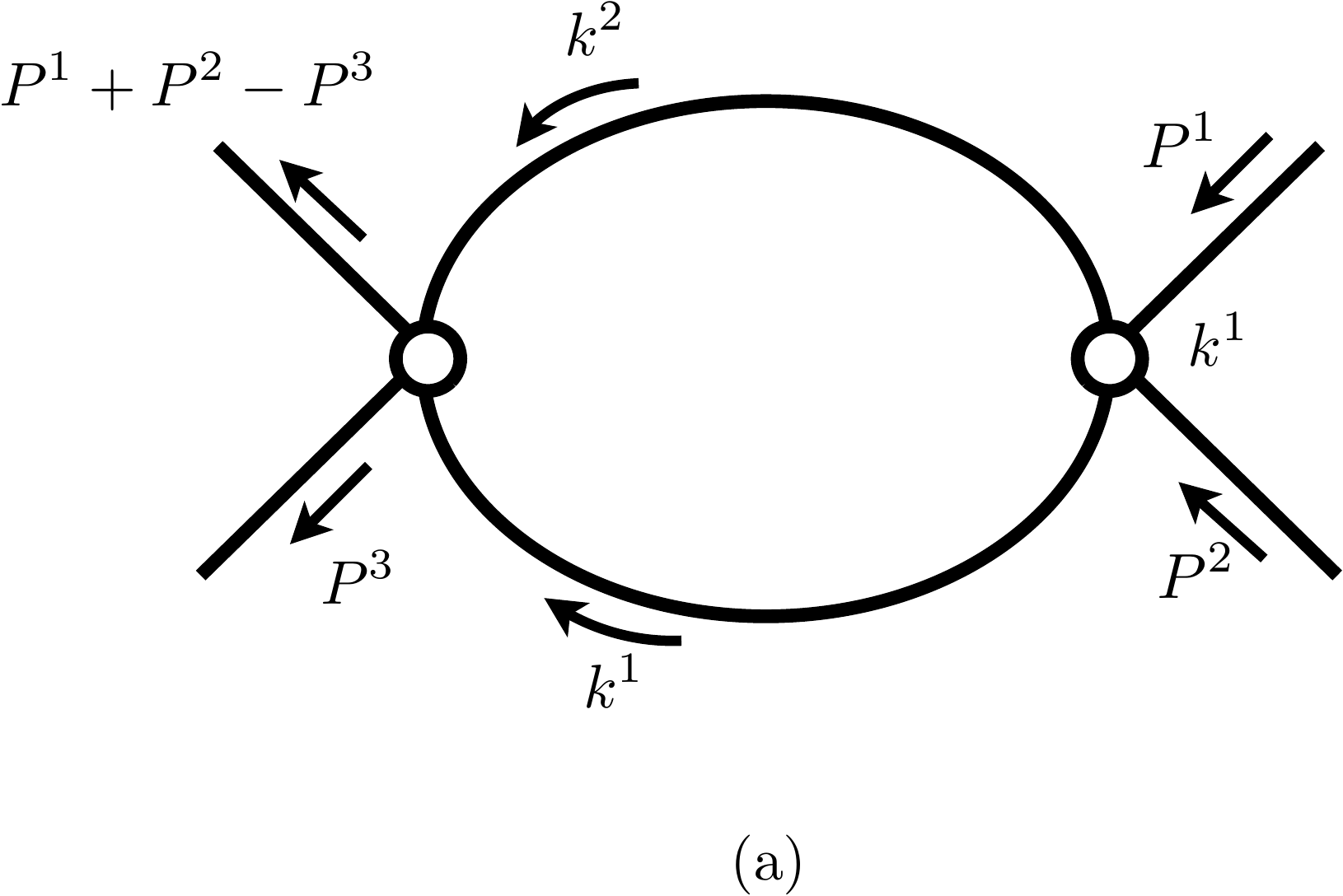}
    \hspace{0.1\textwidth}
    \includegraphics[width=0.3\linewidth]{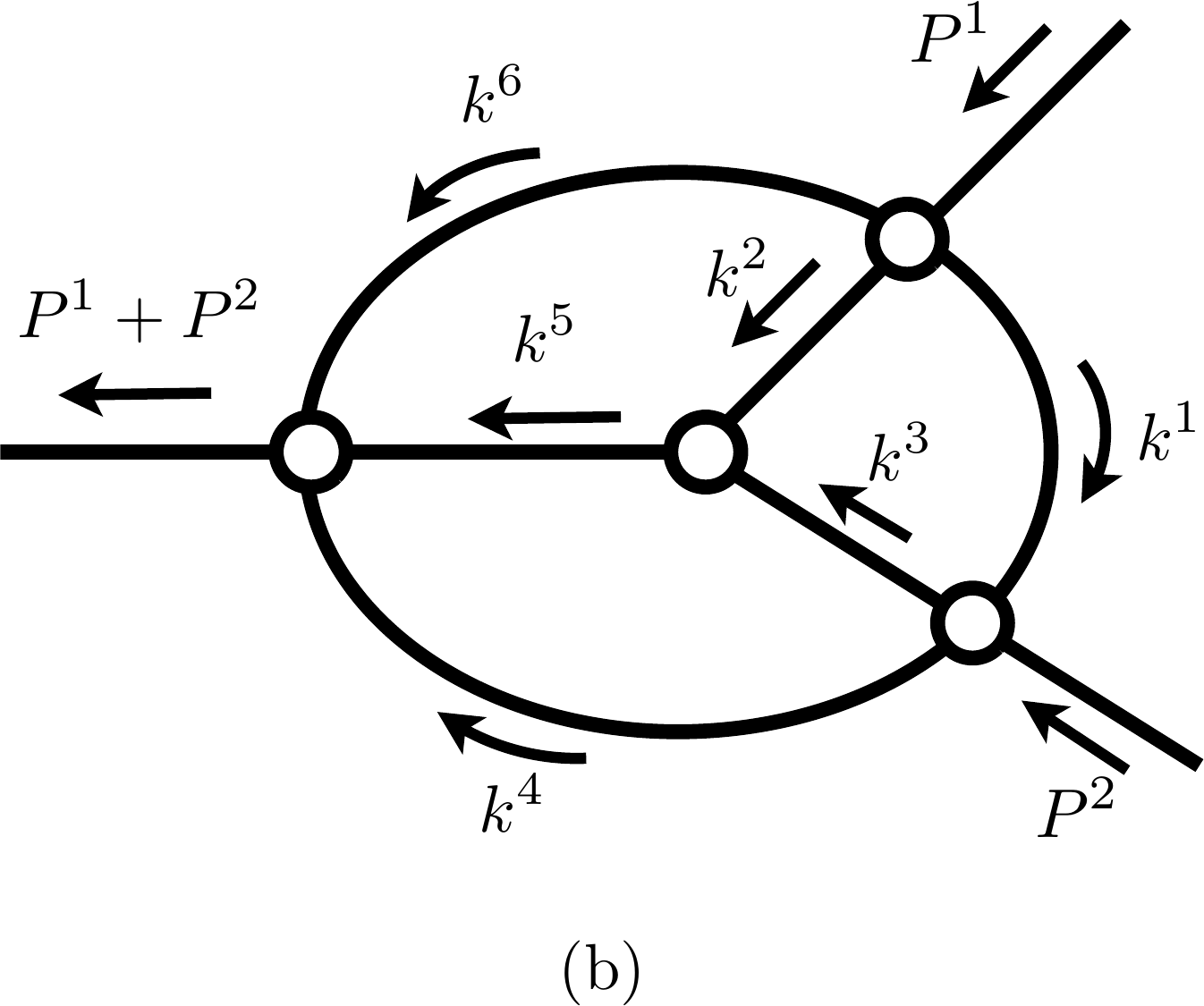}
    \caption{
    (a) For this bubble diagram we can write $k^1 = \ell^1$ and $k^2 = -\ell^1 + P^1 + P^2$.
    (b) For this scarab diagram we can write $k^1 = \ell^1$, $k^4 = \ell^2$, $k^6 = \ell^3$, $k^2 = -\ell^1 - \ell^3 + P^1$, $k^3 = \ell^1 - \ell^2 + P^2$ and $k^5 = -\ell^2 - \ell^3 + P^1 + P^2$.}
    \label{fig:diagrams}
\end{figure*}

Let us consider a $D=1+d$-dimensional quantum field theory in a finite spatial volume $L^d$ with a metric of signature $(+,-,\cdots,-)$.
To preserve translation symmetry, which will ensure that spatial momentum is a good quantum number, we will further impose periodic boundary conditions on this space.
We consider a Feynman diagram with independent external momenta $P^A$, $A = 1,\dots, N_E$, internal momenta $k^i$, $i = 1,\dots, N_P$, and loop momenta $\ell^a$, $a = 1,\dots, N_L$.
Such a diagram has $N_V = N_P - N_L + 1$ vertices and $N_E + 1$ external legs.
Simple examples are given in \cref{fig:diagrams}.

In a finite volume, a generic diagram can be expressed using the usual Feynman rules, as long as one replaces all integrals over the spatial components of the loop momenta with discrete sums.
Alternatively, we will find it more useful to write them in the form
\begin{equation}\label{eq:PoissonSummation}
   iI(P)=\sum_{\vb{n}} iI_{\vb{n}}(P)
\end{equation}
where $P:=(P^A)_{A=1}^{N_E}$ represents the set of all external momenta and
\begin{equation}\label{eq:iIn}
    \begin{aligned}
        iI_{\vb{n}}(P)&:= i^{N_V}\int\ddbar[N_LD]{\ell} e^{-i Ln_a\cdot\ell^a}\prod_{i} D(k_i^2),\\
        D(k^2)&:=\frac{i}{k^2 - m^2 + i\epsilon},
    \end{aligned}
\end{equation}
which can be obtained through the Poisson summation formula.
The sum is over all integer $d$-vectors $\mathbf{n}:=(\vb{n}_a)_{a=1}^{N_L}$, for which we also write $n_a = (0,\vb{n}_a)$. 
The integral on the other hand is over all $\ell^a\in\mathbb{R}^{1,d}$ with the integration measure
\begin{equation}
    \ddbar[N_LD]{\ell} := \prod_{a}\frac{\dd[D]{\ell^a}}{(2\pi)^{D}}.
\end{equation}
The advantage of the formulation used here is that the integral corresponding to setting all the $\vb{n}_a=\vb{0}$ is the infinite volume diagram, so that finite volume corrections are given by the sum in  \cref{eq:PoissonSummation} with $\vb{n}\neq\vb{0}$. 
Although the $\epsilon$-prescription appearing in the propagator is not the same as the $\varepsilon$ in \cref{eq:TM_gen}, it also pushes the spectrum of the theory into the complex plane and thus plays an analogous role.

\textbf{\textit{Boost-Averaging.}}---Next, we show how boost-averaging can be used to suppress these finite-volume corrections.
To do this, we need to isolate the Lorentz-breaking effects on this Feynman diagram. 
This can be done by using the traditional tools of Feynman parametrization~\cite{Coleman:2018mew, Weinzierl:2022eaz}, in terms of which 
\begin{equation}\label{eq:feynman_params}
    \begin{aligned}
        iI_{\vb{n}}(P) &=(N_P-1)! \int_{[0,1]^{N_P}}\dd[N_P]{u} \delta\qty(\sum_i u^i - 1)\\
        &\hphantom{{}={}}\times e^{i L n_a \cdot p^a} iJ_{\vb{n}}(u,P),\\
        iJ_{\vb{n}}(u,P) &:= i^{N_V}\int\ddbar[N_LD]{\ell} e^{-i L n_a\cdot\ell^a}\qty(\frac{i}{g_{ab}\ell^a\cdot \ell^b - \Delta})^{N_P}.
\end{aligned}
\end{equation}
The integrand is given in terms of the topology of the diagram, which is codified by the linear relations $k^i = f^i_a \ell^a + F^i_A P^A$.
These allow us to describe the diagram in terms of three elements:
\begin{enumerate}
    \item A metric on the space of loop momenta $g_{ab} = \sum_i u^i f^i_a f^i_b$, whose inverse we will denote by $g^{ab}$ and which will serve us to raise and lower loop indices indicated by lower-case Latin letters. 
    \item A rescaling of the external momenta $p_a = \sum_i u^i f^i_a F^i_A P^A$.
    \item
    An energy-squared scale
    \begin{equation}\label{eq:Delta}
        \begin{aligned}
            \Delta = m^2 + g^{ab}p_a\cdot p_b - \sum_i u^i(F^i_AP^A)^2 - i\epsilon.
        \end{aligned}
    \end{equation}

\end{enumerate}

The advantage of this rewriting is that, due to the absence of terms coupling the loop and external momenta, the integral $J_{\vb{n}}$ is Lorentz invariant.
Thus, the only term that breaks Lorentz symmetry is the exponential $e^{i L n_a\cdot p^a}$.
The technique of boost-averaging consists of choosing a set of external momenta $P$ in the finite-volume theory that are all close to the kinematic point under consideration.
Denoting by $\ev{\bullet}$ averages of quantities with respect to this choice, we obtain that
\begin{equation}
    \begin{aligned}
        \ev{I_{\vb{n}}} &=(N_P-1)! \int_{[0,1]^{N_P}}\dd[N_P]{u} \delta\qty(\sum_i u^i - 1)\\
        &\hphantom{{}={}}\times \ev{e^{i L n_a\cdot p^a}} J_{\vb{n}}(u,P).\\
\end{aligned}
\end{equation}
Given that $|\ev{e^{iLn_a\cdot p^a}}|\leq 1$, this shows that boost averaging indeed produces a suppression of the finite-volume effects.
This is due to the destructive interference that becomes more evident when a large number of equivalent kinematics are considered in the average.

\textbf{\textit{Kinematic Suppression.}}---We now turn to the issue of showing that these finite-volume corrections are, even without boost-averaging, exponentially suppressed for non-zero $\epsilon$.
We will conclude that each kinematic point in the finite-volume theory converges exponentially to the Lorentz invariant infinite-volume result.
The rate of this suppression is controlled by $L\Re\sqrt{\Delta}$ and is faster in kinematic regions for which $\Re\Delta > 0$, so that $\Re\sqrt{\Delta}\approx\sqrt{\Re\Delta}$.
Conversely, for $\Re\Delta < 0$, we have $\Re\sqrt{\Delta} \approx \epsilon/\sqrt{-\Re\Delta}$, so the suppression is much slower and boost-averaging is needed to better approximate the infinite-volume result in practice.

To show this suppression, we follow the standard argument of Feynman parametrization in which one rewrites the loop integrals in a coordinate system $q^\alpha$ determined by a basis $e_\alpha^a$ which is orthogonal with respect to $g_{ab}$.
In this coordinate system $\ell^a = q^\alpha e_\alpha^a$ and the integral becomes
\begin{equation}\label{eq:loop_orthogonalization}
    iJ_{\vb{n}}(u,P) = i^{N_V}\int\frac{\ddbar[N_LD]{q}}{g^{D/2}} e^{-iL e_\alpha^a n_a\cdot q^\alpha} \qty(\frac{i}{q^2 - \Delta})^{N_P},
\end{equation}
where $g = \det g_{\bullet\bullet}$ is the first Symanzik polynomial~\footnote{The reader might find it useful to note that $g\Delta$ is the second Symanzik polynomial.}.
This integral can now be Wick rotated.
Schwinger parametrization then turns it into a Gaussian that can be solved exactly.
The remaining integral over the Schwinger parameter can then be recognized as a representation of the modified Bessel function of the second kind so that
\begin{equation}\label{eq:eps_suppression}
    \begin{aligned}
        J_{\vb{n}}(u,P) &= \frac{1}{(4\pi)^{N_LD/2}g^{D/2}(N_P-1)!}\\
                        &\hphantom{{}={}}\times 2\qty(\frac{2\sqrt{\Delta}}{L\|n\|})^{\nu} K_\nu\qty(L\|n\|\sqrt{\Delta}).
\end{aligned}
\end{equation}
with $\nu = -N_P + N_L D/2$, which is half the energy dimension of the diagram, and $\|n\| = \sqrt{-g^{ab}n_a\cdot n_b}$.
The exponential suppression claimed is then a direct consequence of the asymptotic behavior of the Bessel function for large arguments $K_\nu(z)\sim e^{-z}$. 
Note that to ensure this asymptotic behavior, $L\,\Re\sqrt{\Delta}\gg 1$, which can be achieved for arbitrary kinematics by requiring $L\epsilon/m\gg 1$.
In other words, if one aims to achieve an energy resolution scale of order of $\mathcal{O}(\epsilon)$, one may need volumes satisfying $L\gg m/\epsilon$.

\begin{figure}
    \centering
    \includegraphics[width=0.9\linewidth]{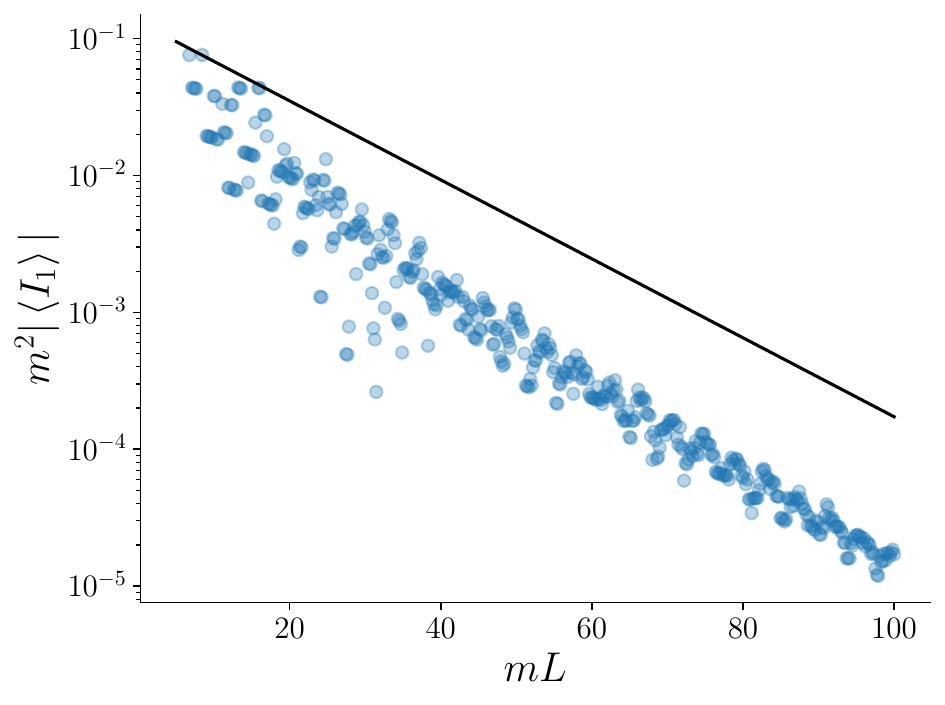}
    \caption{
    Dependence on the volume of $I_1$ for the bubble diagram \cref{fig:diagrams}(a).
    Computed with $D=1+1$, $s = (P^1 + P^2)^2 = (2.5m)^2$ and $\epsilon = 0.1m^2$.
    The solid line corresponds to \cref{eq:final_formula} without averaging and in the center of momentum frame.
    The blue points correspond to the result with averaging by considering all possible kinematics with external momenta lower than $(2\pi/a)/8$ with $ma = 0.1$. Variations of $s$ are allowed with a relative tolerance of $1\%$.
        }
    \label{fig:high_suppression}
\end{figure}

\textbf{\textit{Conclusion and outlook.}}---Putting everything together, we conclude that with our regulators, the finite-volume corrections to a Feynman diagram are given by the sum over $\vb{n}\neq 0$ of
\begin{equation}\label{eq:final_formula}
    \begin{aligned}
        \ev{I_{\vb{n}}} &= \frac{2}{(4\pi)^{N_LD/2}}\int_{[0,1]^{N_P}}\dd[N_P]{u}\delta\qty(\sum_i u^i - 1) \times\\
        &\hphantom{{}={}}g^{-D/2}\ev{e^{iLn_a\cdot p^a}}\qty(\frac{2\sqrt{\Delta}}{L\|n\|})^\nu K_\nu(L\|n\|\sqrt{\Delta}),
    \end{aligned}
\end{equation}
up to derivatives with respect to $n_a$.
This identifies the two main mechanisms of suppression of finite-volume effects, namely, the exponential suppression due to the asymptotic expansion of the Bessel function and the further suppression due to the characteristic function of the boost averaging.
The first gives a proof that the RESOs approach, in principle, captures universally arbitrary scattering observables.
The second shows how one can, in principle, improve the results from this proposal (as well as the wavefunction proposal) systematically.

This immediately informs qualitative aspects of finite-volume errors in employing the RESOs and wavepacket approaches.
For example, in the wavepacket approach, a potential issue is that the reduction of finite-volume effects requires tighter wavepackets, which, in turn, requires broader averages over finite-volume kinematics.
This reduces the resolution of the infinite-volume kinematics.
In the RESOs approach, this is analogous to how one may ``naively'' reduce the size of the finite-volume errors by carelessly increasing the size of the bins considered in the boost averaging.
Of course, experience with RESOs then shows that, in the wavepacket approach, one can tackle this issue by not only attempting to reduce finite-volume errors by tightening the wavepackets, but also by considering the wavepackets scattering in different frames.

A more quantitative feature we can remark is that the finite-volume errors scale as $L^{-\nu-1/2}$.
Therefore, they decrease as the dimension and the number of loops increase.
On the other hand, they increase with the number of external probes. This is consistent with, for example, formalism that has been developed to study electroweak reactions from finite-volume matrix elements~\cite{Davoudi:2020xdv, Baroni:2018iau,Briceno:2015tza, Briceno:2012yi, Lozano:2022kfz}, which have large finite-volume effects.
As we detail in the End Matter, our results have potential applications for lattice QCD calculations on traditional computers as well.

We can further use \cref{eq:final_formula} as a basis for numerical evaluations of finite-volume corrections.
An example of this is shown in \cref{fig:high_suppression}.
We would like to stress here that the graph-theoretic aspects of these integrals have already been thoroughly explored~\cite{Mizera:2021icv}.
Thus, they can be set up efficiently in any dimension.
However, these diagrams will further include non-perturbative vertices, which will have to be estimated on a case-by-case basis for different reactions.

The new control we have acquired over finite-volume errors greatly increases our confidence in the new avenues quantum computers are offering for the calculation of scattering observables through lattice QCD.
They further show that these calculations are systematically improvable, aligning with the ab initio spirit of lattice QCD as a whole.
Furthermore, for the RESOs approach, this finishes the proof that this proposal gives a rigorous and universal framework for the determination of scattering observables.
This has no counterpart in the methods that have been developed for traditional computers.
We are thus left with a note of encouragement and the need for developing an explicit implementation of this proposal on quantum hardware. 
\section*{Acknowledgments}

We acknowledge Christian W. Bauer, Fernando Romero-L\'opez, Martin J. Savage, and Stephen R. Sharpe for useful comments on the manuscript. IMB, ANC and RAB were supported by the U.S. Department of Energy, Office of Science under Award No. DE-AC02-05CH11231.
IMB and ANC were additionally supported by U.S. Department of Energy, Office of Science, National Quantum Information Science Research Centers, Quantum Systems Accelerator and Quantum Information Science Enabled Discovery (QuantISED) for High Energy Physics (KA2401032). 
IMB was further supported in part by the Lawrence Berkeley National Lab LDRD project No. LDRD25-131. MAC acknowledges support from the U.S. Department of Energy Contract No. DE-AC05-06OR23177, under which Jefferson Science Associates, LLC, manages and operates Jefferson Lab.

\appendix

\section{Generalization to non-scalar theories}
\label{app:generalization}

 None of our arguments used to arrive at our main result depended on the vectors $n_a$ being spatial.
Thus, by taking derivatives with respect to $n_a$, we can multiply the integrand of \cref{eq:iIn} by arbitrary polynomials of the loop momenta.
This allows us to lift the assumption that we are working with a scalar field theory and shows that our arguments are general.

For example, let us consider the case of a vector interaction. In such a case, we will in general, have to consider loops of the form, 
\begin{equation}
    \begin{aligned}
   iK^{a}_{\mu}(P)&:= \sum_{\vb{n}} 
   i^{N_V}\int\ddbar[N_LD]{\ell} e^{-i Ln_a\cdot\ell^a}\ell^a_\mu\prod_{i} D(k_i^2)\\
   &= \frac{i}{L}\sum_{\vb{n}}\pdv{n_a^\mu}iI_{\vb{n}}(P)
    \end{aligned}
\end{equation}
which we have written as a partial derivative of our functions, $I_{\vb{n}}(P)$. Consequently, since we know the asymptotic behavior of $I_{\vb{n}}(P)$, we claim we can obtain the asymptotic behavior of $iK^{a}_{\mu}(P)$ and any other finite-volume function. In particular, since the derivative of Bessel functions is still a Bessel function, the suppression in \eqref{eq:final_formula} holds for arbitrary theories as well.

\section{Boost averaging details}

\begin{figure}
    \centering
    \includegraphics[width=0.9\linewidth]{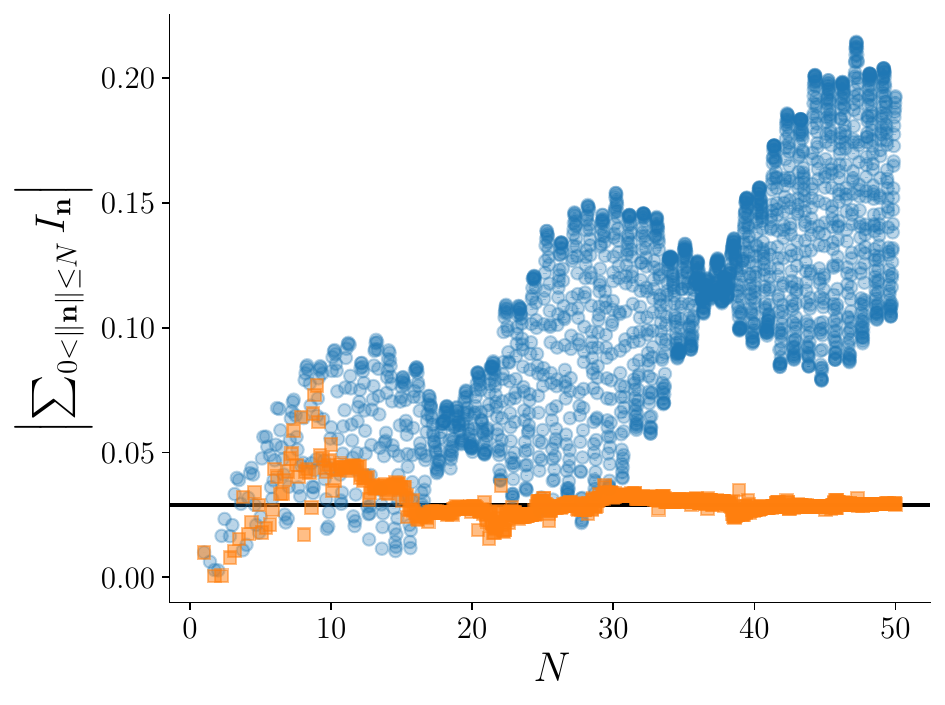}[t]
    \caption{Convergence of \cref{eq:PoissonSummation} for the center of momentum frame of the bubble diagram of \cref{fig:diagrams}(a). In here $D = 1 + 3$, $s=(P^1 + P^2)^2 = (2.4m)^2$, $mL = 10$ and $\epsilon = 0$.
    The blue circles show the partial sums, which we order as a sum over shells of increasing radii. 
    For the orange squares, we first turn this sum into a power series, with the coefficient of degree $n$ corresponding to the $n$-th shell. 
    The orange boxes then correspond to the diagonal terms in the associated Pad\'e table.
    The black line shows the result obtained through traditional methods (see, for example, Ref.~\cite{Kim:2005gf}, where it was labeled $F_s$).  
    }   \label{fig:pade}
\end{figure}

Boost averaging can be done through some probability measure on the space of external momenta with a given set of kinematic invariants.
Then the expectation value of this exponential term can be identified with the characteristic function of this measure.
One can get a rough idea of the magnitude of this suppression for large values of $L$ by considering the cumulant expansion
\begin{equation}\label{eq:boost_avg2}
    \ev{e^{iLn_a\cdot p^a}}=e^{iLn_a\cdot\ev{p^a} - \frac{1}{2}L^2\sigma^2 + \cdots},
\end{equation}
with $\sigma^2=\langle(n_a\cdot p^a)^2\rangle-\langle n_a\cdot p^a\rangle^2$.
In here, the ellipsis contains higher cumulants.
For Gaussian averaging (such as the ones considered in the wavepacket approach), these higher cumulants vanish, and thus, this suppression is exponential. 

In practice, this boost-averaging cannot be done without considering a nontrivial window of different kinematic invariants in the infinite-volume limit.
This yields an error in the kinematic point under consideration.
This error can be reduced at the expense of reducing the number of finite-volume kinematic points available, and thus the $\sigma^2$ controlling this exponential suppression.
This poses an optimization problem that, as far as we know, must be solved numerically on a case-by-case basis. We point the reader to Refs.~\cite{Briceno:2020rar, Carrillo:2024chu} for proof-of-principle investigations in this problem and its resolution in practice. 
We also note the possibility of employing twisted boundary conditions \cite{Briceno:2013hya} to increase the number of kinematic points available with the same infinite-volume limit.
We will leave explorations along this direction to future work.

\section{Impact to traditional lattice calculations}

We now turn to the possibility of performing the sum in \cref{eq:PoissonSummation}.
Given that the suppression rate is proportional to $\|n\|$, this can be reasonably approximated with a finite number of terms.
However, since it is also proportional to $\epsilon$, the smaller we make this regulator, the more terms we must include.
In fact, at $\epsilon = 0$, this series is divergent.
For example, it is known that, in $D = 1+1$, the finite-volume corrections to the bubble diagram of \cref{fig:diagrams}(a) constitute a geometric series.
This suggests that we should try to sum \cref{eq:PoissonSummation} in more general cases using Padé approximants.
We give a first indication that this can be a promising method in \cref{fig:pade}.

This last remark suggests that our results may also impact future calculations in lattice QCD performed on traditional computers.
In regards to scattering, these calculations rely heavily on the calculation of \cref{eq:PoissonSummation} at $\epsilon = 0$ for the bubble and triangle diagrams present in, for example, the aforementioned formalism for two-body matrix elements~\cite{Davoudi:2020xdv, Baroni:2018iau,Briceno:2015tza, Briceno:2012yi}. Ultimately, these formalisms require the value of these functions in the $\epsilon = 0$ limit, where this series does not converge. Consequently, these formalisms generally do not utilize the Poison summation formula. Instead, the integrals over the zero components of the momenta are evaluated analytically, and the spatial components are left to be evaluated either analytically, as in the case of the bubble diagram, or numerically, as in the triangle diagram~\cite{Baroni:2018iau}. This procedure is particularly slow for the triangle diagram. 
The combination of \cref{eq:final_formula} along with Pad\'e summation could help speed this class of calculations. 
We will leave this exploration for future studies. 


\begin{thebibliography}{94}%
\makeatletter
\providecommand \@ifxundefined [1]{%
 \@ifx{#1\undefined}
}%
\providecommand \@ifnum [1]{%
 \ifnum #1\expandafter \@firstoftwo
 \else \expandafter \@secondoftwo
 \fi
}%
\providecommand \@ifx [1]{%
 \ifx #1\expandafter \@firstoftwo
 \else \expandafter \@secondoftwo
 \fi
}%
\providecommand \natexlab [1]{#1}%
\providecommand \enquote  [1]{``#1''}%
\providecommand \bibnamefont  [1]{#1}%
\providecommand \bibfnamefont [1]{#1}%
\providecommand \citenamefont [1]{#1}%
\providecommand \href@noop [0]{\@secondoftwo}%
\providecommand \href [0]{\begingroup \@sanitize@url \@href}%
\providecommand \@href[1]{\@@startlink{#1}\@@href}%
\providecommand \@@href[1]{\endgroup#1\@@endlink}%
\providecommand \@sanitize@url [0]{\catcode `\\12\catcode `\$12\catcode `\&12\catcode `\#12\catcode `\^12\catcode `\_12\catcode `\%12\relax}%
\providecommand \@@startlink[1]{}%
\providecommand \@@endlink[0]{}%
\providecommand \url  [0]{\begingroup\@sanitize@url \@url }%
\providecommand \@url [1]{\endgroup\@href {#1}{\urlprefix }}%
\providecommand \urlprefix  [0]{URL }%
\providecommand \Eprint [0]{\href }%
\providecommand \doibase [0]{https://doi.org/}%
\providecommand \selectlanguage [0]{\@gobble}%
\providecommand \bibinfo  [0]{\@secondoftwo}%
\providecommand \bibfield  [0]{\@secondoftwo}%
\providecommand \translation [1]{[#1]}%
\providecommand \BibitemOpen [0]{}%
\providecommand \bibitemStop [0]{}%
\providecommand \bibitemNoStop [0]{.\EOS\space}%
\providecommand \EOS [0]{\spacefactor3000\relax}%
\providecommand \BibitemShut  [1]{\csname bibitem#1\endcsname}%
\let\auto@bib@innerbib\@empty
\bibitem [{\citenamefont {Acciarri}\ \emph {et~al.}(2016)\citenamefont {Acciarri} \emph {et~al.}}]{DUNE:2016hlj}%
  \BibitemOpen
  \bibfield  {author} {\bibinfo {author} {\bibfnamefont {R.}~\bibnamefont {Acciarri}} \emph {et~al.} (\bibinfo {collaboration} {DUNE}),\ }\href@noop {} {\bibinfo {title} {{Long-Baseline Neutrino Facility (LBNF) and Deep Underground Neutrino Experiment (DUNE)}: {Conceptual Design Report, Volume 1: The LBNF and DUNE Projects}}} (\bibinfo {year} {2016}),\ \Eprint {https://arxiv.org/abs/1601.05471} {arXiv:1601.05471 [physics.ins-det]} \BibitemShut {NoStop}%
\bibitem [{\citenamefont {Abi}\ \emph {et~al.}(2021)\citenamefont {Abi} \emph {et~al.}}]{DUNE:2020fgq}%
  \BibitemOpen
  \bibfield  {author} {\bibinfo {author} {\bibfnamefont {B.}~\bibnamefont {Abi}} \emph {et~al.} (\bibinfo {collaboration} {DUNE}),\ }\bibfield  {title} {\bibinfo {title} {{Prospects for beyond the Standard Model physics searches at the Deep Underground Neutrino Experiment}},\ }\href {https://doi.org/10.1140/epjc/s10052-021-09007-w} {\bibfield  {journal} {\bibinfo  {journal} {Eur. Phys. J. C}\ }\textbf {\bibinfo {volume} {81}},\ \bibinfo {pages} {322} (\bibinfo {year} {2021})},\ \Eprint {https://arxiv.org/abs/2008.12769} {arXiv:2008.12769 [hep-ex]} \BibitemShut {NoStop}%
\bibitem [{\citenamefont {Accardi}\ \emph {et~al.}(2016)\citenamefont {Accardi} \emph {et~al.}}]{Accardi:2012qut}%
  \BibitemOpen
  \bibfield  {author} {\bibinfo {author} {\bibfnamefont {A.}~\bibnamefont {Accardi}} \emph {et~al.},\ }\bibfield  {title} {\bibinfo {title} {{Electron Ion Collider: The Next QCD Frontier}: {Understanding the glue that binds us all}},\ }\href {https://doi.org/10.1140/epja/i2016-16268-9} {\bibfield  {journal} {\bibinfo  {journal} {Eur. Phys. J. A}\ }\textbf {\bibinfo {volume} {52}},\ \bibinfo {pages} {268} (\bibinfo {year} {2016})},\ \Eprint {https://arxiv.org/abs/1212.1701} {arXiv:1212.1701 [nucl-ex]} \BibitemShut {NoStop}%
\bibitem [{\citenamefont {Abdul~Khalek}\ \emph {et~al.}(2022)\citenamefont {Abdul~Khalek} \emph {et~al.}}]{AbdulKhalek:2022hcn}%
  \BibitemOpen
  \bibfield  {author} {\bibinfo {author} {\bibfnamefont {R.}~\bibnamefont {Abdul~Khalek}} \emph {et~al.},\ }\href@noop {} {\bibinfo {title} {{Snowmass 2021 White Paper: Electron Ion Collider for High Energy Physics}}} (\bibinfo {year} {2022}),\ \Eprint {https://arxiv.org/abs/2203.13199} {arXiv:2203.13199 [hep-ph]} \BibitemShut {NoStop}%
\bibitem [{\citenamefont {Luscher}(1986{\natexlab{a}})}]{Luscher:1985dn}%
  \BibitemOpen
  \bibfield  {author} {\bibinfo {author} {\bibfnamefont {M.}~\bibnamefont {Luscher}},\ }\bibfield  {title} {\bibinfo {title} {{Volume Dependence of the Energy Spectrum in Massive Quantum Field Theories. 1. Stable Particle States}},\ }\href {https://doi.org/10.1007/BF01211589} {\bibfield  {journal} {\bibinfo  {journal} {Commun. Math. Phys.}\ }\textbf {\bibinfo {volume} {104}},\ \bibinfo {pages} {177} (\bibinfo {year} {1986}{\natexlab{a}})}\BibitemShut {NoStop}%
\bibitem [{\citenamefont {Luscher}(1986{\natexlab{b}})}]{Luscher:1986pf}%
  \BibitemOpen
  \bibfield  {author} {\bibinfo {author} {\bibfnamefont {M.}~\bibnamefont {Luscher}},\ }\bibfield  {title} {\bibinfo {title} {{Volume Dependence of the Energy Spectrum in Massive Quantum Field Theories. 2. Scattering States}},\ }\href {https://doi.org/10.1007/BF01211097} {\bibfield  {journal} {\bibinfo  {journal} {Commun. Math. Phys.}\ }\textbf {\bibinfo {volume} {105}},\ \bibinfo {pages} {153} (\bibinfo {year} {1986}{\natexlab{b}})}\BibitemShut {NoStop}%
\bibitem [{\citenamefont {Luscher}(1991)}]{Luscher:1990ux}%
  \BibitemOpen
  \bibfield  {author} {\bibinfo {author} {\bibfnamefont {M.}~\bibnamefont {Luscher}},\ }\bibfield  {title} {\bibinfo {title} {{Two particle states on a torus and their relation to the scattering matrix}},\ }\href {https://doi.org/10.1016/0550-3213(91)90366-6} {\bibfield  {journal} {\bibinfo  {journal} {Nucl. Phys. B}\ }\textbf {\bibinfo {volume} {354}},\ \bibinfo {pages} {531} (\bibinfo {year} {1991})}\BibitemShut {NoStop}%
\bibitem [{\citenamefont {Briceno}\ \emph {et~al.}(2018)\citenamefont {Briceno}, \citenamefont {Dudek},\ and\ \citenamefont {Young}}]{Briceno:2017max}%
  \BibitemOpen
  \bibfield  {author} {\bibinfo {author} {\bibfnamefont {R.~A.}\ \bibnamefont {Briceno}}, \bibinfo {author} {\bibfnamefont {J.~J.}\ \bibnamefont {Dudek}},\ and\ \bibinfo {author} {\bibfnamefont {R.~D.}\ \bibnamefont {Young}},\ }\bibfield  {title} {\bibinfo {title} {{Scattering processes and resonances from lattice QCD}},\ }\href {https://doi.org/10.1103/RevModPhys.90.025001} {\bibfield  {journal} {\bibinfo  {journal} {Rev. Mod. Phys.}\ }\textbf {\bibinfo {volume} {90}},\ \bibinfo {pages} {025001} (\bibinfo {year} {2018})},\ \Eprint {https://arxiv.org/abs/1706.06223} {arXiv:1706.06223} \BibitemShut {NoStop}%
\bibitem [{\citenamefont {Hansen}\ and\ \citenamefont {Sharpe}(2019)}]{Hansen:2019nir}%
  \BibitemOpen
  \bibfield  {author} {\bibinfo {author} {\bibfnamefont {M.~T.}\ \bibnamefont {Hansen}}\ and\ \bibinfo {author} {\bibfnamefont {S.~R.}\ \bibnamefont {Sharpe}},\ }\bibfield  {title} {\bibinfo {title} {{Lattice QCD and Three-particle Decays of Resonances}},\ }\href {https://doi.org/10.1146/annurev-nucl-101918-023723} {\bibfield  {journal} {\bibinfo  {journal} {Ann. Rev. Nucl. Part. Sci.}\ }\textbf {\bibinfo {volume} {69}},\ \bibinfo {pages} {65} (\bibinfo {year} {2019})},\ \Eprint {https://arxiv.org/abs/1901.00483} {arXiv:1901.00483 [hep-lat]} \BibitemShut {NoStop}%
\bibitem [{\citenamefont {Mai}\ \emph {et~al.}(2021)\citenamefont {Mai}, \citenamefont {D\"oring},\ and\ \citenamefont {Rusetsky}}]{Mai:2021lwb}%
  \BibitemOpen
  \bibfield  {author} {\bibinfo {author} {\bibfnamefont {M.}~\bibnamefont {Mai}}, \bibinfo {author} {\bibfnamefont {M.}~\bibnamefont {D\"oring}},\ and\ \bibinfo {author} {\bibfnamefont {A.}~\bibnamefont {Rusetsky}},\ }\bibfield  {title} {\bibinfo {title} {{Multi-particle systems on the lattice and chiral extrapolations: a brief review}},\ }\href {https://doi.org/10.1140/epjs/s11734-021-00146-5} {\bibfield  {journal} {\bibinfo  {journal} {Eur. Phys. J. ST}\ }\textbf {\bibinfo {volume} {230}},\ \bibinfo {pages} {1623} (\bibinfo {year} {2021})},\ \Eprint {https://arxiv.org/abs/2103.00577} {arXiv:2103.00577 [hep-lat]} \BibitemShut {NoStop}%
\bibitem [{\citenamefont {Bulava}\ \emph {et~al.}(2022)\citenamefont {Bulava} \emph {et~al.}}]{Bulava:2022ovd}%
  \BibitemOpen
  \bibfield  {author} {\bibinfo {author} {\bibfnamefont {J.}~\bibnamefont {Bulava}} \emph {et~al.},\ }\bibfield  {title} {\bibinfo {title} {{Hadron Spectroscopy with Lattice QCD}},\ }in\ \href@noop {} {\emph {\bibinfo {booktitle} {{Snowmass 2021}}}}\ (\bibinfo {year} {2022})\ \Eprint {https://arxiv.org/abs/2203.03230} {arXiv:2203.03230 [hep-lat]} \BibitemShut {NoStop}%
\bibitem [{Note1()}]{Note1}%
  \BibitemOpen
  \bibinfo {note} {Which happens when the energy of the system is equal to or greater than the sum of the masses of the three particles}\BibitemShut {NoStop}%
\bibitem [{\citenamefont {Dawid}\ \emph {et~al.}(2025)\citenamefont {Dawid}, \citenamefont {Draper}, \citenamefont {Hanlon}, \citenamefont {H\"orz}, \citenamefont {Morningstar}, \citenamefont {Romero-L\'opez}, \citenamefont {Sharpe},\ and\ \citenamefont {Skinner}}]{Dawid:2025zxc}%
  \BibitemOpen
  \bibfield  {author} {\bibinfo {author} {\bibfnamefont {S.~M.}\ \bibnamefont {Dawid}}, \bibinfo {author} {\bibfnamefont {Z.~T.}\ \bibnamefont {Draper}}, \bibinfo {author} {\bibfnamefont {A.~D.}\ \bibnamefont {Hanlon}}, \bibinfo {author} {\bibfnamefont {B.}~\bibnamefont {H\"orz}}, \bibinfo {author} {\bibfnamefont {C.}~\bibnamefont {Morningstar}}, \bibinfo {author} {\bibfnamefont {F.}~\bibnamefont {Romero-L\'opez}}, \bibinfo {author} {\bibfnamefont {S.~R.}\ \bibnamefont {Sharpe}},\ and\ \bibinfo {author} {\bibfnamefont {S.}~\bibnamefont {Skinner}},\ }\href@noop {} {\bibinfo {title} {{QCD predictions for physical multimeson scattering amplitudes}}} (\bibinfo {year} {2025}),\ \Eprint {https://arxiv.org/abs/2502.14348} {arXiv:2502.14348 [hep-lat]} \BibitemShut {NoStop}%
\bibitem [{\citenamefont {Jackura}(2023)}]{Jackura:2022gib}%
  \BibitemOpen
  \bibfield  {author} {\bibinfo {author} {\bibfnamefont {A.~W.}\ \bibnamefont {Jackura}},\ }\bibfield  {title} {\bibinfo {title} {{Three-body scattering and quantization conditions from S-matrix unitarity}},\ }\href {https://doi.org/10.1103/PhysRevD.108.034505} {\bibfield  {journal} {\bibinfo  {journal} {Phys. Rev. D}\ }\textbf {\bibinfo {volume} {108}},\ \bibinfo {pages} {034505} (\bibinfo {year} {2023})},\ \Eprint {https://arxiv.org/abs/2208.10587} {arXiv:2208.10587 [hep-lat]} \BibitemShut {NoStop}%
\bibitem [{\citenamefont {Hansen}\ \emph {et~al.}(2021)\citenamefont {Hansen}, \citenamefont {Brice\~no}, \citenamefont {Edwards}, \citenamefont {Thomas},\ and\ \citenamefont {Wilson}}]{Hansen:2020otl}%
  \BibitemOpen
  \bibfield  {author} {\bibinfo {author} {\bibfnamefont {M.~T.}\ \bibnamefont {Hansen}}, \bibinfo {author} {\bibfnamefont {R.~A.}\ \bibnamefont {Brice\~no}}, \bibinfo {author} {\bibfnamefont {R.~G.}\ \bibnamefont {Edwards}}, \bibinfo {author} {\bibfnamefont {C.~E.}\ \bibnamefont {Thomas}},\ and\ \bibinfo {author} {\bibfnamefont {D.~J.}\ \bibnamefont {Wilson}} (\bibinfo {collaboration} {Hadron Spectrum}),\ }\bibfield  {title} {\bibinfo {title} {{Energy-Dependent $\pi^+ \pi^+ \pi^+$ Scattering Amplitude from QCD}},\ }\href {https://doi.org/10.1103/PhysRevLett.126.012001} {\bibfield  {journal} {\bibinfo  {journal} {Phys. Rev. Lett.}\ }\textbf {\bibinfo {volume} {126}},\ \bibinfo {pages} {012001} (\bibinfo {year} {2021})},\ \Eprint {https://arxiv.org/abs/2009.04931} {arXiv:2009.04931 [hep-lat]} \BibitemShut {NoStop}%
\bibitem [{\citenamefont {Brice\~no}\ \emph {et~al.}(2025)\citenamefont {Brice\~no}, \citenamefont {Costa},\ and\ \citenamefont {Jackura}}]{Briceno:2024ehy}%
  \BibitemOpen
  \bibfield  {author} {\bibinfo {author} {\bibfnamefont {R.~A.}\ \bibnamefont {Brice\~no}}, \bibinfo {author} {\bibfnamefont {C.~S.~R.}\ \bibnamefont {Costa}},\ and\ \bibinfo {author} {\bibfnamefont {A.~W.}\ \bibnamefont {Jackura}},\ }\bibfield  {title} {\bibinfo {title} {{Partial-wave projection of relativistic three-body amplitudes}},\ }\href {https://doi.org/10.1103/PhysRevD.111.036029} {\bibfield  {journal} {\bibinfo  {journal} {Phys. Rev. D}\ }\textbf {\bibinfo {volume} {111}},\ \bibinfo {pages} {036029} (\bibinfo {year} {2025})},\ \Eprint {https://arxiv.org/abs/2409.15577} {arXiv:2409.15577 [hep-ph]} \BibitemShut {NoStop}%
\bibitem [{\citenamefont {Mai}\ and\ \citenamefont {Doring}(2019)}]{Mai:2018djl}%
  \BibitemOpen
  \bibfield  {author} {\bibinfo {author} {\bibfnamefont {M.}~\bibnamefont {Mai}}\ and\ \bibinfo {author} {\bibfnamefont {M.}~\bibnamefont {Doring}},\ }\bibfield  {title} {\bibinfo {title} {{Finite-Volume Spectrum of $\pi^+\pi^+$ and $\pi^+\pi^+\pi^+$ Systems}},\ }\href {https://doi.org/10.1103/PhysRevLett.122.062503} {\bibfield  {journal} {\bibinfo  {journal} {Phys. Rev. Lett.}\ }\textbf {\bibinfo {volume} {122}},\ \bibinfo {pages} {062503} (\bibinfo {year} {2019})},\ \Eprint {https://arxiv.org/abs/1807.04746} {arXiv:1807.04746 [hep-lat]} \BibitemShut {NoStop}%
\bibitem [{\citenamefont {Mai}\ and\ \citenamefont {D\"oring}(2017)}]{Mai:2017bge}%
  \BibitemOpen
  \bibfield  {author} {\bibinfo {author} {\bibfnamefont {M.}~\bibnamefont {Mai}}\ and\ \bibinfo {author} {\bibfnamefont {M.}~\bibnamefont {D\"oring}},\ }\bibfield  {title} {\bibinfo {title} {{Three-body Unitarity in the Finite Volume}},\ }\href {https://doi.org/10.1140/epja/i2017-12440-1} {\bibfield  {journal} {\bibinfo  {journal} {Eur. Phys. J. A}\ }\textbf {\bibinfo {volume} {53}},\ \bibinfo {pages} {240} (\bibinfo {year} {2017})},\ \Eprint {https://arxiv.org/abs/1709.08222} {arXiv:1709.08222 [hep-lat]} \BibitemShut {NoStop}%
\bibitem [{\citenamefont {Jordan}\ \emph {et~al.}(2018)\citenamefont {Jordan}, \citenamefont {Krovi}, \citenamefont {Lee},\ and\ \citenamefont {Preskill}}]{Jordan:2017lea}%
  \BibitemOpen
  \bibfield  {author} {\bibinfo {author} {\bibfnamefont {S.~P.}\ \bibnamefont {Jordan}}, \bibinfo {author} {\bibfnamefont {H.}~\bibnamefont {Krovi}}, \bibinfo {author} {\bibfnamefont {K.~S.~M.}\ \bibnamefont {Lee}},\ and\ \bibinfo {author} {\bibfnamefont {J.}~\bibnamefont {Preskill}},\ }\bibfield  {title} {\bibinfo {title} {{BQP-completeness of Scattering in Scalar Quantum Field Theory}},\ }\href {https://doi.org/10.22331/q-2018-01-08-44} {\bibfield  {journal} {\bibinfo  {journal} {Quantum}\ }\textbf {\bibinfo {volume} {2}},\ \bibinfo {pages} {44} (\bibinfo {year} {2018})},\ \Eprint {https://arxiv.org/abs/1703.00454} {arXiv:1703.00454 [quant-ph]} \BibitemShut {NoStop}%
\bibitem [{\citenamefont {Brice\~no}\ \emph {et~al.}(2021)\citenamefont {Brice\~no}, \citenamefont {Guerrero}, \citenamefont {Hansen},\ and\ \citenamefont {Sturzu}}]{Briceno:2020rar}%
  \BibitemOpen
  \bibfield  {author} {\bibinfo {author} {\bibfnamefont {R.~A.}\ \bibnamefont {Brice\~no}}, \bibinfo {author} {\bibfnamefont {J.~V.}\ \bibnamefont {Guerrero}}, \bibinfo {author} {\bibfnamefont {M.~T.}\ \bibnamefont {Hansen}},\ and\ \bibinfo {author} {\bibfnamefont {A.~M.}\ \bibnamefont {Sturzu}},\ }\bibfield  {title} {\bibinfo {title} {{Role of boundary conditions in quantum computations of scattering observables}},\ }\href {https://doi.org/10.1103/PhysRevD.103.014506} {\bibfield  {journal} {\bibinfo  {journal} {Phys. Rev. D}\ }\textbf {\bibinfo {volume} {103}},\ \bibinfo {pages} {014506} (\bibinfo {year} {2021})},\ \Eprint {https://arxiv.org/abs/2007.01155} {arXiv:2007.01155 [hep-lat]} \BibitemShut {NoStop}%
\bibitem [{\citenamefont {Brice\~no}\ \emph {et~al.}(2022)\citenamefont {Brice\~no}, \citenamefont {Carrillo}, \citenamefont {Guerrero}, \citenamefont {Hansen},\ and\ \citenamefont {Sturzu}}]{Briceno:2021aiw}%
  \BibitemOpen
  \bibfield  {author} {\bibinfo {author} {\bibfnamefont {R.~A.}\ \bibnamefont {Brice\~no}}, \bibinfo {author} {\bibfnamefont {M.~A.}\ \bibnamefont {Carrillo}}, \bibinfo {author} {\bibfnamefont {J.~V.}\ \bibnamefont {Guerrero}}, \bibinfo {author} {\bibfnamefont {M.~T.}\ \bibnamefont {Hansen}},\ and\ \bibinfo {author} {\bibfnamefont {A.~M.}\ \bibnamefont {Sturzu}},\ }\bibfield  {title} {\bibinfo {title} {{Accessing scattering amplitudes using quantum computers}},\ }\href {https://doi.org/10.22323/1.396.0315} {\bibfield  {journal} {\bibinfo  {journal} {PoS}\ }\textbf {\bibinfo {volume} {LATTICE2021}},\ \bibinfo {pages} {315} (\bibinfo {year} {2022})},\ \Eprint {https://arxiv.org/abs/2112.01968} {arXiv:2112.01968 [hep-lat]} \BibitemShut {NoStop}%
\bibitem [{\citenamefont {Carrillo}\ \emph {et~al.}(2024)\citenamefont {Carrillo}, \citenamefont {Brice\~no},\ and\ \citenamefont {Sturzu}}]{Carrillo:2024chu}%
  \BibitemOpen
  \bibfield  {author} {\bibinfo {author} {\bibfnamefont {M.~A.}\ \bibnamefont {Carrillo}}, \bibinfo {author} {\bibfnamefont {R.~A.}\ \bibnamefont {Brice\~no}},\ and\ \bibinfo {author} {\bibfnamefont {A.~M.}\ \bibnamefont {Sturzu}},\ }\bibfield  {title} {\bibinfo {title} {{Inclusive reactions from finite Minkowski spacetime correlation functions}},\ }\href {https://doi.org/10.1103/PhysRevD.110.054503} {\bibfield  {journal} {\bibinfo  {journal} {Phys. Rev. D}\ }\textbf {\bibinfo {volume} {110}},\ \bibinfo {pages} {054503} (\bibinfo {year} {2024})},\ \Eprint {https://arxiv.org/abs/2406.06877} {arXiv:2406.06877 [hep-lat]} \BibitemShut {NoStop}%
\bibitem [{\citenamefont {Raychowdhury}\ and\ \citenamefont {Stryker}(2020)}]{Raychowdhury:2019iki}%
  \BibitemOpen
  \bibfield  {author} {\bibinfo {author} {\bibfnamefont {I.}~\bibnamefont {Raychowdhury}}\ and\ \bibinfo {author} {\bibfnamefont {J.~R.}\ \bibnamefont {Stryker}},\ }\bibfield  {title} {\bibinfo {title} {{Loop, string, and hadron dynamics in SU(2) Hamiltonian lattice gauge theories}},\ }\href {https://doi.org/10.1103/PhysRevD.101.114502} {\bibfield  {journal} {\bibinfo  {journal} {Phys. Rev. D}\ }\textbf {\bibinfo {volume} {101}},\ \bibinfo {pages} {114502} (\bibinfo {year} {2020})},\ \Eprint {https://arxiv.org/abs/1912.06133} {arXiv:1912.06133 [hep-lat]} \BibitemShut {NoStop}%
\bibitem [{\citenamefont {Kadam}\ \emph {et~al.}(2023)\citenamefont {Kadam}, \citenamefont {Raychowdhury},\ and\ \citenamefont {Stryker}}]{Kadam:2022ipf}%
  \BibitemOpen
  \bibfield  {author} {\bibinfo {author} {\bibfnamefont {S.~V.}\ \bibnamefont {Kadam}}, \bibinfo {author} {\bibfnamefont {I.}~\bibnamefont {Raychowdhury}},\ and\ \bibinfo {author} {\bibfnamefont {J.~R.}\ \bibnamefont {Stryker}},\ }\bibfield  {title} {\bibinfo {title} {{Loop-string-hadron formulation of an SU(3) gauge theory with dynamical quarks}},\ }\href {https://doi.org/10.1103/PhysRevD.107.094513} {\bibfield  {journal} {\bibinfo  {journal} {Phys. Rev. D}\ }\textbf {\bibinfo {volume} {107}},\ \bibinfo {pages} {094513} (\bibinfo {year} {2023})},\ \Eprint {https://arxiv.org/abs/2212.04490} {arXiv:2212.04490 [hep-lat]} \BibitemShut {NoStop}%
\bibitem [{\citenamefont {Kadam}\ \emph {et~al.}(2025)\citenamefont {Kadam}, \citenamefont {Naskar}, \citenamefont {Raychowdhury},\ and\ \citenamefont {Stryker}}]{Kadam:2024ifg}%
  \BibitemOpen
  \bibfield  {author} {\bibinfo {author} {\bibfnamefont {S.~V.}\ \bibnamefont {Kadam}}, \bibinfo {author} {\bibfnamefont {A.}~\bibnamefont {Naskar}}, \bibinfo {author} {\bibfnamefont {I.}~\bibnamefont {Raychowdhury}},\ and\ \bibinfo {author} {\bibfnamefont {J.~R.}\ \bibnamefont {Stryker}},\ }\bibfield  {title} {\bibinfo {title} {{Loop-string-hadron approach to SU(3) lattice Yang-Mills theory: Hilbert space of a trivalent vertex}},\ }\href {https://doi.org/10.1103/PhysRevD.111.074516} {\bibfield  {journal} {\bibinfo  {journal} {Phys. Rev. D}\ }\textbf {\bibinfo {volume} {111}},\ \bibinfo {pages} {074516} (\bibinfo {year} {2025})},\ \Eprint {https://arxiv.org/abs/2407.19181} {arXiv:2407.19181 [hep-lat]} \BibitemShut {NoStop}%
\bibitem [{\citenamefont {Ciavarella}\ \emph {et~al.}(2021)\citenamefont {Ciavarella}, \citenamefont {Klco},\ and\ \citenamefont {Savage}}]{Ciavarella:2021nmj}%
  \BibitemOpen
  \bibfield  {author} {\bibinfo {author} {\bibfnamefont {A.}~\bibnamefont {Ciavarella}}, \bibinfo {author} {\bibfnamefont {N.}~\bibnamefont {Klco}},\ and\ \bibinfo {author} {\bibfnamefont {M.~J.}\ \bibnamefont {Savage}},\ }\bibfield  {title} {\bibinfo {title} {{Trailhead for quantum simulation of SU(3) Yang-Mills lattice gauge theory in the local multiplet basis}},\ }\href {https://doi.org/10.1103/PhysRevD.103.094501} {\bibfield  {journal} {\bibinfo  {journal} {Phys. Rev. D}\ }\textbf {\bibinfo {volume} {103}},\ \bibinfo {pages} {094501} (\bibinfo {year} {2021})},\ \Eprint {https://arxiv.org/abs/2101.10227} {arXiv:2101.10227 [quant-ph]} \BibitemShut {NoStop}%
\bibitem [{\citenamefont {D'Andrea}\ \emph {et~al.}(2024)\citenamefont {D'Andrea}, \citenamefont {Bauer}, \citenamefont {Grabowska},\ and\ \citenamefont {Freytsis}}]{DAndrea:2023qnr}%
  \BibitemOpen
  \bibfield  {author} {\bibinfo {author} {\bibfnamefont {I.}~\bibnamefont {D'Andrea}}, \bibinfo {author} {\bibfnamefont {C.~W.}\ \bibnamefont {Bauer}}, \bibinfo {author} {\bibfnamefont {D.~M.}\ \bibnamefont {Grabowska}},\ and\ \bibinfo {author} {\bibfnamefont {M.}~\bibnamefont {Freytsis}},\ }\bibfield  {title} {\bibinfo {title} {{New basis for Hamiltonian SU(2) simulations}},\ }\href {https://doi.org/10.1103/PhysRevD.109.074501} {\bibfield  {journal} {\bibinfo  {journal} {Phys. Rev. D}\ }\textbf {\bibinfo {volume} {109}},\ \bibinfo {pages} {074501} (\bibinfo {year} {2024})},\ \Eprint {https://arxiv.org/abs/2307.11829} {arXiv:2307.11829 [hep-ph]} \BibitemShut {NoStop}%
\bibitem [{\citenamefont {Grabowska}\ \emph {et~al.}(2025)\citenamefont {Grabowska}, \citenamefont {Kane},\ and\ \citenamefont {Bauer}}]{Grabowska:2024emw}%
  \BibitemOpen
  \bibfield  {author} {\bibinfo {author} {\bibfnamefont {D.~M.}\ \bibnamefont {Grabowska}}, \bibinfo {author} {\bibfnamefont {C.~F.}\ \bibnamefont {Kane}},\ and\ \bibinfo {author} {\bibfnamefont {C.~W.}\ \bibnamefont {Bauer}},\ }\bibfield  {title} {\bibinfo {title} {{Fully gauge-fixed SU(2) Hamiltonian for quantum simulations}},\ }\href {https://doi.org/10.1103/PhysRevD.111.114516} {\bibfield  {journal} {\bibinfo  {journal} {Phys. Rev. D}\ }\textbf {\bibinfo {volume} {111}},\ \bibinfo {pages} {114516} (\bibinfo {year} {2025})},\ \Eprint {https://arxiv.org/abs/2409.10610} {arXiv:2409.10610 [quant-ph]} \BibitemShut {NoStop}%
\bibitem [{\citenamefont {Burbano}\ and\ \citenamefont {Bauer}(2025)}]{Burbano:2024uvn}%
  \BibitemOpen
  \bibfield  {author} {\bibinfo {author} {\bibfnamefont {I.~M.}\ \bibnamefont {Burbano}}\ and\ \bibinfo {author} {\bibfnamefont {C.~W.}\ \bibnamefont {Bauer}},\ }\bibfield  {title} {\bibinfo {title} {{Gauge loop-string-hadron formulation on general graphs and applications to fully gauge fixed Hamiltonian lattice gauge theory}},\ }\href {https://doi.org/10.1007/JHEP12(2025)060} {\bibfield  {journal} {\bibinfo  {journal} {JHEP}\ }\textbf {\bibinfo {volume} {12}},\ \bibinfo {pages} {060}},\ \Eprint {https://arxiv.org/abs/2409.13812} {arXiv:2409.13812 [hep-lat]} \BibitemShut {NoStop}%
\bibitem [{\citenamefont {Bergner}\ \emph {et~al.}(2024)\citenamefont {Bergner}, \citenamefont {Hanada}, \citenamefont {Rinaldi},\ and\ \citenamefont {Schafer}}]{Bergner:2024qjl}%
  \BibitemOpen
  \bibfield  {author} {\bibinfo {author} {\bibfnamefont {G.}~\bibnamefont {Bergner}}, \bibinfo {author} {\bibfnamefont {M.}~\bibnamefont {Hanada}}, \bibinfo {author} {\bibfnamefont {E.}~\bibnamefont {Rinaldi}},\ and\ \bibinfo {author} {\bibfnamefont {A.}~\bibnamefont {Schafer}},\ }\bibfield  {title} {\bibinfo {title} {{Toward QCD on quantum computer: orbifold lattice approach}},\ }\href {https://doi.org/10.1007/JHEP05(2024)234} {\bibfield  {journal} {\bibinfo  {journal} {JHEP}\ }\textbf {\bibinfo {volume} {05}},\ \bibinfo {pages} {234}},\ \Eprint {https://arxiv.org/abs/2401.12045} {arXiv:2401.12045 [hep-th]} \BibitemShut {NoStop}%
\bibitem [{\citenamefont {Ciavarella}\ and\ \citenamefont {Bauer}(2024)}]{Ciavarella:2024fzw}%
  \BibitemOpen
  \bibfield  {author} {\bibinfo {author} {\bibfnamefont {A.~N.}\ \bibnamefont {Ciavarella}}\ and\ \bibinfo {author} {\bibfnamefont {C.~W.}\ \bibnamefont {Bauer}},\ }\bibfield  {title} {\bibinfo {title} {{Quantum Simulation of SU(3) Lattice Yang-Mills Theory at Leading Order in Large-Nc Expansion}},\ }\href {https://doi.org/10.1103/PhysRevLett.133.111901} {\bibfield  {journal} {\bibinfo  {journal} {Phys. Rev. Lett.}\ }\textbf {\bibinfo {volume} {133}},\ \bibinfo {pages} {111901} (\bibinfo {year} {2024})},\ \Eprint {https://arxiv.org/abs/2402.10265} {arXiv:2402.10265 [hep-ph]} \BibitemShut {NoStop}%
\bibitem [{\citenamefont {Ciavarella}\ \emph {et~al.}(2025)\citenamefont {Ciavarella}, \citenamefont {Burbano},\ and\ \citenamefont {Bauer}}]{Ciavarella:2025bsg}%
  \BibitemOpen
  \bibfield  {author} {\bibinfo {author} {\bibfnamefont {A.~N.}\ \bibnamefont {Ciavarella}}, \bibinfo {author} {\bibfnamefont {I.~M.}\ \bibnamefont {Burbano}},\ and\ \bibinfo {author} {\bibfnamefont {C.~W.}\ \bibnamefont {Bauer}},\ }\bibfield  {title} {\bibinfo {title} {{Efficient truncations of SU(Nc) lattice gauge theory for quantum simulation}},\ }\href {https://doi.org/10.1103/ylqb-phv5} {\bibfield  {journal} {\bibinfo  {journal} {Phys. Rev. D}\ }\textbf {\bibinfo {volume} {112}},\ \bibinfo {pages} {054514} (\bibinfo {year} {2025})},\ \Eprint {https://arxiv.org/abs/2503.11888} {arXiv:2503.11888 [hep-lat]} \BibitemShut {NoStop}%
\bibitem [{\citenamefont {Assi}\ and\ \citenamefont {Lamm}(2024)}]{Assi:2024pdn}%
  \BibitemOpen
  \bibfield  {author} {\bibinfo {author} {\bibfnamefont {B.}~\bibnamefont {Assi}}\ and\ \bibinfo {author} {\bibfnamefont {H.}~\bibnamefont {Lamm}},\ }\bibfield  {title} {\bibinfo {title} {{Digitization and subduction of SU(N) gauge theories}},\ }\href {https://doi.org/10.1103/PhysRevD.110.074511} {\bibfield  {journal} {\bibinfo  {journal} {Phys. Rev. D}\ }\textbf {\bibinfo {volume} {110}},\ \bibinfo {pages} {074511} (\bibinfo {year} {2024})},\ \Eprint {https://arxiv.org/abs/2405.12204} {arXiv:2405.12204 [hep-lat]} \BibitemShut {NoStop}%
\bibitem [{\citenamefont {Charles}\ \emph {et~al.}(2024)\citenamefont {Charles}, \citenamefont {Gustafson}, \citenamefont {Hardt}, \citenamefont {Herren}, \citenamefont {Hogan}, \citenamefont {Lamm}, \citenamefont {Starecheski}, \citenamefont {Van~de Water},\ and\ \citenamefont {Wagman}}]{Charles:2023zbl}%
  \BibitemOpen
  \bibfield  {author} {\bibinfo {author} {\bibfnamefont {C.}~\bibnamefont {Charles}}, \bibinfo {author} {\bibfnamefont {E.~J.}\ \bibnamefont {Gustafson}}, \bibinfo {author} {\bibfnamefont {E.}~\bibnamefont {Hardt}}, \bibinfo {author} {\bibfnamefont {F.}~\bibnamefont {Herren}}, \bibinfo {author} {\bibfnamefont {N.}~\bibnamefont {Hogan}}, \bibinfo {author} {\bibfnamefont {H.}~\bibnamefont {Lamm}}, \bibinfo {author} {\bibfnamefont {S.}~\bibnamefont {Starecheski}}, \bibinfo {author} {\bibfnamefont {R.~S.}\ \bibnamefont {Van~de Water}},\ and\ \bibinfo {author} {\bibfnamefont {M.~L.}\ \bibnamefont {Wagman}},\ }\bibfield  {title} {\bibinfo {title} {{Simulating Z2 lattice gauge theory on a quantum computer}},\ }\href {https://doi.org/10.1103/PhysRevE.109.015307} {\bibfield  {journal} {\bibinfo  {journal} {Phys. Rev. E}\ }\textbf {\bibinfo {volume} {109}},\ \bibinfo {pages} {015307} (\bibinfo {year} {2024})},\ \Eprint {https://arxiv.org/abs/2305.02361} {arXiv:2305.02361 [hep-lat]} \BibitemShut {NoStop}%
\bibitem [{\citenamefont {Bennewitz}\ \emph {et~al.}(2025)\citenamefont {Bennewitz} \emph {et~al.}}]{Bennewitz:2025nhz}%
  \BibitemOpen
  \bibfield  {author} {\bibinfo {author} {\bibfnamefont {E.~R.}\ \bibnamefont {Bennewitz}} \emph {et~al.},\ }\bibfield  {title} {\bibinfo {title} {{Simulating Meson Scattering on Spin Quantum Simulators}},\ }\href {https://doi.org/10.22331/q-2025-06-17-1773} {\bibfield  {journal} {\bibinfo  {journal} {Quantum}\ }\textbf {\bibinfo {volume} {9}},\ \bibinfo {pages} {1773} (\bibinfo {year} {2025})},\ \Eprint {https://arxiv.org/abs/2403.07061} {arXiv:2403.07061 [quant-ph]} \BibitemShut {NoStop}%
\bibitem [{\citenamefont {Davoudi}\ \emph {et~al.}(2025)\citenamefont {Davoudi}, \citenamefont {Hsieh},\ and\ \citenamefont {Kadam}}]{Davoudi:2025rdv}%
  \BibitemOpen
  \bibfield  {author} {\bibinfo {author} {\bibfnamefont {Z.}~\bibnamefont {Davoudi}}, \bibinfo {author} {\bibfnamefont {C.-C.}\ \bibnamefont {Hsieh}},\ and\ \bibinfo {author} {\bibfnamefont {S.~V.}\ \bibnamefont {Kadam}},\ }\bibfield  {title} {\bibinfo {title} {{Quantum computation of hadron scattering in a lattice gauge theory}},\ }\href@noop {} {\  (\bibinfo {year} {2025})},\ \Eprint {https://arxiv.org/abs/2505.20408} {arXiv:2505.20408 [quant-ph]} \BibitemShut {NoStop}%
\bibitem [{\citenamefont {Schuhmacher}\ \emph {et~al.}(2025)\citenamefont {Schuhmacher}, \citenamefont {Su}, \citenamefont {Osborne}, \citenamefont {Gandon}, \citenamefont {Halimeh},\ and\ \citenamefont {Tavernelli}}]{Schuhmacher:2025ehh}%
  \BibitemOpen
  \bibfield  {author} {\bibinfo {author} {\bibfnamefont {J.}~\bibnamefont {Schuhmacher}}, \bibinfo {author} {\bibfnamefont {G.-X.}\ \bibnamefont {Su}}, \bibinfo {author} {\bibfnamefont {J.~J.}\ \bibnamefont {Osborne}}, \bibinfo {author} {\bibfnamefont {A.}~\bibnamefont {Gandon}}, \bibinfo {author} {\bibfnamefont {J.~C.}\ \bibnamefont {Halimeh}},\ and\ \bibinfo {author} {\bibfnamefont {I.}~\bibnamefont {Tavernelli}},\ }\bibfield  {title} {\bibinfo {title} {{Observation of hadron scattering in a lattice gauge theory on a quantum computer}},\ }\href@noop {} {\  (\bibinfo {year} {2025})},\ \Eprint {https://arxiv.org/abs/2505.20387} {arXiv:2505.20387 [quant-ph]} \BibitemShut {NoStop}%
\bibitem [{\citenamefont {Farrell}\ \emph {et~al.}(2025{\natexlab{a}})\citenamefont {Farrell}, \citenamefont {Zemlevskiy}, \citenamefont {Illa},\ and\ \citenamefont {Preskill}}]{Farrell:2025nkx}%
  \BibitemOpen
  \bibfield  {author} {\bibinfo {author} {\bibfnamefont {R.~C.}\ \bibnamefont {Farrell}}, \bibinfo {author} {\bibfnamefont {N.~A.}\ \bibnamefont {Zemlevskiy}}, \bibinfo {author} {\bibfnamefont {M.}~\bibnamefont {Illa}},\ and\ \bibinfo {author} {\bibfnamefont {J.}~\bibnamefont {Preskill}},\ }\bibfield  {title} {\bibinfo {title} {{Digital quantum simulations of scattering in quantum field theories using W states}},\ }\href@noop {} {\  (\bibinfo {year} {2025}{\natexlab{a}})},\ \Eprint {https://arxiv.org/abs/2505.03111} {arXiv:2505.03111 [quant-ph]} \BibitemShut {NoStop}%
\bibitem [{\citenamefont {Ingoldby}\ \emph {et~al.}(2025)\citenamefont {Ingoldby}, \citenamefont {Spannowsky}, \citenamefont {Sypchenko}, \citenamefont {Williams},\ and\ \citenamefont {Wingate}}]{Ingoldby:2025bdb}%
  \BibitemOpen
  \bibfield  {author} {\bibinfo {author} {\bibfnamefont {J.}~\bibnamefont {Ingoldby}}, \bibinfo {author} {\bibfnamefont {M.}~\bibnamefont {Spannowsky}}, \bibinfo {author} {\bibfnamefont {T.}~\bibnamefont {Sypchenko}}, \bibinfo {author} {\bibfnamefont {S.}~\bibnamefont {Williams}},\ and\ \bibinfo {author} {\bibfnamefont {M.}~\bibnamefont {Wingate}},\ }\href@noop {} {\bibinfo {title} {{Real-Time Scattering on Quantum Computers via Hamiltonian Truncation}}} (\bibinfo {year} {2025}),\ \Eprint {https://arxiv.org/abs/2505.03878} {arXiv:2505.03878 [quant-ph]} \BibitemShut {NoStop}%
\bibitem [{\citenamefont {Yusf}\ \emph {et~al.}(2025)\citenamefont {Yusf}, \citenamefont {Gan}, \citenamefont {Moffat},\ and\ \citenamefont {Rupak}}]{Yusf:2024igb}%
  \BibitemOpen
  \bibfield  {author} {\bibinfo {author} {\bibfnamefont {M.}~\bibnamefont {Yusf}}, \bibinfo {author} {\bibfnamefont {L.}~\bibnamefont {Gan}}, \bibinfo {author} {\bibfnamefont {C.}~\bibnamefont {Moffat}},\ and\ \bibinfo {author} {\bibfnamefont {G.}~\bibnamefont {Rupak}},\ }\bibfield  {title} {\bibinfo {title} {{Elastic scattering on a quantum computer}},\ }\href {https://doi.org/10.1103/PhysRevC.111.034001} {\bibfield  {journal} {\bibinfo  {journal} {Phys. Rev. C}\ }\textbf {\bibinfo {volume} {111}},\ \bibinfo {pages} {034001} (\bibinfo {year} {2025})},\ \Eprint {https://arxiv.org/abs/2406.09231} {arXiv:2406.09231 [nucl-th]} \BibitemShut {NoStop}%
\bibitem [{\citenamefont {Chai}\ \emph {et~al.}(2025{\natexlab{a}})\citenamefont {Chai}, \citenamefont {Guo},\ and\ \citenamefont {K\"uhn}}]{Chai:2025qhf}%
  \BibitemOpen
  \bibfield  {author} {\bibinfo {author} {\bibfnamefont {Y.}~\bibnamefont {Chai}}, \bibinfo {author} {\bibfnamefont {Y.}~\bibnamefont {Guo}},\ and\ \bibinfo {author} {\bibfnamefont {S.}~\bibnamefont {K\"uhn}},\ }\bibfield  {title} {\bibinfo {title} {{Towards Quantum Simulation of Meson Scattering in a Z2 Lattice Gauge Theory}},\ }\href@noop {} {\  (\bibinfo {year} {2025}{\natexlab{a}})},\ \Eprint {https://arxiv.org/abs/2505.21240} {arXiv:2505.21240 [quant-ph]} \BibitemShut {NoStop}%
\bibitem [{\citenamefont {Abel}\ \emph {et~al.}(2025)\citenamefont {Abel}, \citenamefont {Spannowsky},\ and\ \citenamefont {Williams}}]{Abel:2025zxb}%
  \BibitemOpen
  \bibfield  {author} {\bibinfo {author} {\bibfnamefont {S.}~\bibnamefont {Abel}}, \bibinfo {author} {\bibfnamefont {M.}~\bibnamefont {Spannowsky}},\ and\ \bibinfo {author} {\bibfnamefont {S.}~\bibnamefont {Williams}},\ }\bibfield  {title} {\bibinfo {title} {{Real-time scattering processes with continuous-variable quantum computers}},\ }\href {https://doi.org/10.1103/q36d-w649} {\bibfield  {journal} {\bibinfo  {journal} {Phys. Rev. A}\ }\textbf {\bibinfo {volume} {112}},\ \bibinfo {pages} {012614} (\bibinfo {year} {2025})},\ \Eprint {https://arxiv.org/abs/2502.01767} {arXiv:2502.01767 [quant-ph]} \BibitemShut {NoStop}%
\bibitem [{\citenamefont {Sharma}\ \emph {et~al.}(2024)\citenamefont {Sharma}, \citenamefont {Papenbrock},\ and\ \citenamefont {Platter}}]{Sharma:2023bqu}%
  \BibitemOpen
  \bibfield  {author} {\bibinfo {author} {\bibfnamefont {S.}~\bibnamefont {Sharma}}, \bibinfo {author} {\bibfnamefont {T.}~\bibnamefont {Papenbrock}},\ and\ \bibinfo {author} {\bibfnamefont {L.}~\bibnamefont {Platter}},\ }\bibfield  {title} {\bibinfo {title} {{Scattering phase shifts from a quantum computer}},\ }\href {https://doi.org/10.1103/PhysRevC.109.L061001} {\bibfield  {journal} {\bibinfo  {journal} {Phys. Rev. C}\ }\textbf {\bibinfo {volume} {109}},\ \bibinfo {pages} {L061001} (\bibinfo {year} {2024})},\ \Eprint {https://arxiv.org/abs/2311.09298} {arXiv:2311.09298 [nucl-th]} \BibitemShut {NoStop}%
\bibitem [{\citenamefont {Wu}\ \emph {et~al.}(2024)\citenamefont {Wu}, \citenamefont {Du}, \citenamefont {Zhao},\ and\ \citenamefont {Vary}}]{Wu:2024adk}%
  \BibitemOpen
  \bibfield  {author} {\bibinfo {author} {\bibfnamefont {S.}~\bibnamefont {Wu}}, \bibinfo {author} {\bibfnamefont {W.}~\bibnamefont {Du}}, \bibinfo {author} {\bibfnamefont {X.}~\bibnamefont {Zhao}},\ and\ \bibinfo {author} {\bibfnamefont {J.~P.}\ \bibnamefont {Vary}},\ }\bibfield  {title} {\bibinfo {title} {{Efficient and precise quantum simulation of ultrarelativistic quark-nucleus scattering}},\ }\href {https://doi.org/10.1103/PhysRevD.110.056044} {\bibfield  {journal} {\bibinfo  {journal} {Phys. Rev. D}\ }\textbf {\bibinfo {volume} {110}},\ \bibinfo {pages} {056044} (\bibinfo {year} {2024})},\ \Eprint {https://arxiv.org/abs/2404.00819} {arXiv:2404.00819 [quant-ph]} \BibitemShut {NoStop}%
\bibitem [{\citenamefont {Rajput}\ \emph {et~al.}(2023)\citenamefont {Rajput}, \citenamefont {Roggero},\ and\ \citenamefont {Wiebe}}]{Rajput:2021trn}%
  \BibitemOpen
  \bibfield  {author} {\bibinfo {author} {\bibfnamefont {A.}~\bibnamefont {Rajput}}, \bibinfo {author} {\bibfnamefont {A.}~\bibnamefont {Roggero}},\ and\ \bibinfo {author} {\bibfnamefont {N.}~\bibnamefont {Wiebe}},\ }\bibfield  {title} {\bibinfo {title} {{Quantum error correction with gauge symmetries}},\ }\href {https://doi.org/10.1038/s41534-023-00706-8} {\bibfield  {journal} {\bibinfo  {journal} {npj Quantum Inf.}\ }\textbf {\bibinfo {volume} {9}},\ \bibinfo {pages} {41} (\bibinfo {year} {2023})},\ \Eprint {https://arxiv.org/abs/2112.05186} {arXiv:2112.05186 [quant-ph]} \BibitemShut {NoStop}%
\bibitem [{\citenamefont {Yao}(2025)}]{Yao:2025cxs}%
  \BibitemOpen
  \bibfield  {author} {\bibinfo {author} {\bibfnamefont {X.}~\bibnamefont {Yao}},\ }\href@noop {} {\bibinfo {title} {{Quantum Error Correction Codes for Truncated SU(2) Lattice Gauge Theories}}} (\bibinfo {year} {2025}),\ \Eprint {https://arxiv.org/abs/2511.13721} {arXiv:2511.13721 [quant-ph]} \BibitemShut {NoStop}%
\bibitem [{\citenamefont {Halimeh}\ \emph {et~al.}(2022{\natexlab{a}})\citenamefont {Halimeh}, \citenamefont {Lang},\ and\ \citenamefont {Hauke}}]{Halimeh:2021vzf}%
  \BibitemOpen
  \bibfield  {author} {\bibinfo {author} {\bibfnamefont {J.~C.}\ \bibnamefont {Halimeh}}, \bibinfo {author} {\bibfnamefont {H.}~\bibnamefont {Lang}},\ and\ \bibinfo {author} {\bibfnamefont {P.}~\bibnamefont {Hauke}},\ }\bibfield  {title} {\bibinfo {title} {{Gauge protection in non-abelian lattice gauge theories}},\ }\href {https://doi.org/10.1088/1367-2630/ac5564} {\bibfield  {journal} {\bibinfo  {journal} {New J. Phys.}\ }\textbf {\bibinfo {volume} {24}},\ \bibinfo {pages} {033015} (\bibinfo {year} {2022}{\natexlab{a}})},\ \Eprint {https://arxiv.org/abs/2106.09032} {arXiv:2106.09032 [cond-mat.quant-gas]} \BibitemShut {NoStop}%
\bibitem [{\citenamefont {Halimeh}\ \emph {et~al.}(2022{\natexlab{b}})\citenamefont {Halimeh}, \citenamefont {Homeier}, \citenamefont {Schweizer}, \citenamefont {Aidelsburger}, \citenamefont {Hauke},\ and\ \citenamefont {Grusdt}}]{Halimeh:2021lnv}%
  \BibitemOpen
  \bibfield  {author} {\bibinfo {author} {\bibfnamefont {J.~C.}\ \bibnamefont {Halimeh}}, \bibinfo {author} {\bibfnamefont {L.}~\bibnamefont {Homeier}}, \bibinfo {author} {\bibfnamefont {C.}~\bibnamefont {Schweizer}}, \bibinfo {author} {\bibfnamefont {M.}~\bibnamefont {Aidelsburger}}, \bibinfo {author} {\bibfnamefont {P.}~\bibnamefont {Hauke}},\ and\ \bibinfo {author} {\bibfnamefont {F.}~\bibnamefont {Grusdt}},\ }\bibfield  {title} {\bibinfo {title} {{Stabilizing lattice gauge theories through simplified local pseudogenerators}},\ }\href {https://doi.org/10.1103/PhysRevResearch.4.033120} {\bibfield  {journal} {\bibinfo  {journal} {Phys. Rev. Res.}\ }\textbf {\bibinfo {volume} {4}},\ \bibinfo {pages} {033120} (\bibinfo {year} {2022}{\natexlab{b}})},\ \Eprint {https://arxiv.org/abs/2108.02203} {arXiv:2108.02203 [cond-mat.quant-gas]} \BibitemShut {NoStop}%
\bibitem [{\citenamefont {Chen}\ \emph {et~al.}(2024)\citenamefont {Chen}, \citenamefont {Gorshkov},\ and\ \citenamefont {Xu}}]{Chen:2022dox}%
  \BibitemOpen
  \bibfield  {author} {\bibinfo {author} {\bibfnamefont {Y.-A.}\ \bibnamefont {Chen}}, \bibinfo {author} {\bibfnamefont {A.~V.}\ \bibnamefont {Gorshkov}},\ and\ \bibinfo {author} {\bibfnamefont {Y.}~\bibnamefont {Xu}},\ }\bibfield  {title} {\bibinfo {title} {{Error-correcting codes for fermionic quantum simulation}},\ }\href {https://doi.org/10.21468/SciPostPhys.16.1.033} {\bibfield  {journal} {\bibinfo  {journal} {SciPost Phys.}\ }\textbf {\bibinfo {volume} {16}},\ \bibinfo {pages} {033} (\bibinfo {year} {2024})},\ \Eprint {https://arxiv.org/abs/2210.08411} {arXiv:2210.08411 [quant-ph]} \BibitemShut {NoStop}%
\bibitem [{\citenamefont {Stryker}(2019)}]{Stryker:2018efp}%
  \BibitemOpen
  \bibfield  {author} {\bibinfo {author} {\bibfnamefont {J.~R.}\ \bibnamefont {Stryker}},\ }\bibfield  {title} {\bibinfo {title} {{Oracles for Gauss's law on digital quantum computers}},\ }\href {https://doi.org/10.1103/PhysRevA.99.042301} {\bibfield  {journal} {\bibinfo  {journal} {Phys. Rev. A}\ }\textbf {\bibinfo {volume} {99}},\ \bibinfo {pages} {042301} (\bibinfo {year} {2019})},\ \Eprint {https://arxiv.org/abs/1812.01617} {arXiv:1812.01617 [quant-ph]} \BibitemShut {NoStop}%
\bibitem [{\citenamefont {Spagnoli}\ \emph {et~al.}(2026)\citenamefont {Spagnoli}, \citenamefont {Roggero},\ and\ \citenamefont {Wiebe}}]{Spagnoli:2024mib}%
  \BibitemOpen
  \bibfield  {author} {\bibinfo {author} {\bibfnamefont {L.}~\bibnamefont {Spagnoli}}, \bibinfo {author} {\bibfnamefont {A.}~\bibnamefont {Roggero}},\ and\ \bibinfo {author} {\bibfnamefont {N.}~\bibnamefont {Wiebe}},\ }\bibfield  {title} {\bibinfo {title} {{Fault-tolerant simulation of Lattice Gauge Theories with gauge covariant codes}},\ }\href {https://doi.org/10.22331/q-2026-01-16-1968} {\bibfield  {journal} {\bibinfo  {journal} {Quantum}\ }\textbf {\bibinfo {volume} {10}},\ \bibinfo {pages} {1968} (\bibinfo {year} {2026})},\ \Eprint {https://arxiv.org/abs/2405.19293} {arXiv:2405.19293 [quant-ph]} \BibitemShut {NoStop}%
\bibitem [{\citenamefont {Carena}\ \emph {et~al.}(2024)\citenamefont {Carena}, \citenamefont {Lamm}, \citenamefont {Li},\ and\ \citenamefont {Liu}}]{Carena:2024dzu}%
  \BibitemOpen
  \bibfield  {author} {\bibinfo {author} {\bibfnamefont {M.}~\bibnamefont {Carena}}, \bibinfo {author} {\bibfnamefont {H.}~\bibnamefont {Lamm}}, \bibinfo {author} {\bibfnamefont {Y.-Y.}\ \bibnamefont {Li}},\ and\ \bibinfo {author} {\bibfnamefont {W.}~\bibnamefont {Liu}},\ }\bibfield  {title} {\bibinfo {title} {{Quantum error thresholds for gauge-redundant digitizations of lattice field theories}},\ }\href {https://doi.org/10.1103/PhysRevD.110.054516} {\bibfield  {journal} {\bibinfo  {journal} {Phys. Rev. D}\ }\textbf {\bibinfo {volume} {110}},\ \bibinfo {pages} {054516} (\bibinfo {year} {2024})},\ \Eprint {https://arxiv.org/abs/2402.16780} {arXiv:2402.16780 [hep-lat]} \BibitemShut {NoStop}%
\bibitem [{\citenamefont {Faist}\ \emph {et~al.}(2020)\citenamefont {Faist}, \citenamefont {Nezami}, \citenamefont {Albert}, \citenamefont {Salton}, \citenamefont {Pastawski}, \citenamefont {Hayden},\ and\ \citenamefont {Preskill}}]{Faist:2019ahr}%
  \BibitemOpen
  \bibfield  {author} {\bibinfo {author} {\bibfnamefont {P.}~\bibnamefont {Faist}}, \bibinfo {author} {\bibfnamefont {S.}~\bibnamefont {Nezami}}, \bibinfo {author} {\bibfnamefont {V.~V.}\ \bibnamefont {Albert}}, \bibinfo {author} {\bibfnamefont {G.}~\bibnamefont {Salton}}, \bibinfo {author} {\bibfnamefont {F.}~\bibnamefont {Pastawski}}, \bibinfo {author} {\bibfnamefont {P.}~\bibnamefont {Hayden}},\ and\ \bibinfo {author} {\bibfnamefont {J.}~\bibnamefont {Preskill}},\ }\bibfield  {title} {\bibinfo {title} {{Continuous symmetries and approximate quantum error correction}},\ }\href {https://doi.org/10.1103/PhysRevX.10.041018} {\bibfield  {journal} {\bibinfo  {journal} {Phys. Rev. X}\ }\textbf {\bibinfo {volume} {10}},\ \bibinfo {pages} {041018} (\bibinfo {year} {2020})},\ \Eprint {https://arxiv.org/abs/1902.07714} {arXiv:1902.07714 [quant-ph]} \BibitemShut {NoStop}%
\bibitem [{\citenamefont {Halimeh}\ \emph {et~al.}(2025)\citenamefont {Halimeh}, \citenamefont {Mueller}, \citenamefont {Knolle}, \citenamefont {Papi{\'c}},\ and\ \citenamefont {Davoudi}}]{Halimeh:2025vvp}%
  \BibitemOpen
  \bibfield  {author} {\bibinfo {author} {\bibfnamefont {J.~C.}\ \bibnamefont {Halimeh}}, \bibinfo {author} {\bibfnamefont {N.}~\bibnamefont {Mueller}}, \bibinfo {author} {\bibfnamefont {J.}~\bibnamefont {Knolle}}, \bibinfo {author} {\bibfnamefont {Z.}~\bibnamefont {Papi{\'c}}},\ and\ \bibinfo {author} {\bibfnamefont {Z.}~\bibnamefont {Davoudi}},\ }\href@noop {} {\bibinfo {title} {{Quantum simulation of out-of-equilibrium dynamics in gauge theories}}} (\bibinfo {year} {2025}),\ \Eprint {https://arxiv.org/abs/2509.03586} {arXiv:2509.03586 [quant-ph]} \BibitemShut {NoStop}%
\bibitem [{\citenamefont {Davoudi}(2025)}]{Davoudi:2025kxb}%
  \BibitemOpen
  \bibfield  {author} {\bibinfo {author} {\bibfnamefont {Z.}~\bibnamefont {Davoudi}},\ }\href@noop {} {\bibinfo {title} {{TASI/CERN/KITP Lecture Notes on ''Toward Quantum Computing Gauge Theories of Nature''}}} (\bibinfo {year} {2025}),\ \Eprint {https://arxiv.org/abs/2507.15840} {arXiv:2507.15840 [hep-lat]} \BibitemShut {NoStop}%
\bibitem [{\citenamefont {Bauer}(2025)}]{Bauer:2025nzf}%
  \BibitemOpen
  \bibfield  {author} {\bibinfo {author} {\bibfnamefont {C.~W.}\ \bibnamefont {Bauer}},\ }\bibfield  {title} {\bibinfo {title} {{Efficient use of quantum computers for collider physics}},\ }\href {https://doi.org/10.1007/JHEP11(2025)108} {\bibfield  {journal} {\bibinfo  {journal} {JHEP}\ }\textbf {\bibinfo {volume} {11}},\ \bibinfo {pages} {108}},\ \Eprint {https://arxiv.org/abs/2503.16602} {arXiv:2503.16602 [hep-ph]} \BibitemShut {NoStop}%
\bibitem [{\citenamefont {Di~Meglio}\ \emph {et~al.}(2024)\citenamefont {Di~Meglio} \emph {et~al.}}]{DiMeglio:2023nsa}%
  \BibitemOpen
  \bibfield  {author} {\bibinfo {author} {\bibfnamefont {A.}~\bibnamefont {Di~Meglio}} \emph {et~al.},\ }\bibfield  {title} {\bibinfo {title} {{Quantum Computing for High-Energy Physics: State of the Art and Challenges}},\ }\href {https://doi.org/10.1103/PRXQuantum.5.037001} {\bibfield  {journal} {\bibinfo  {journal} {PRX Quantum}\ }\textbf {\bibinfo {volume} {5}},\ \bibinfo {pages} {037001} (\bibinfo {year} {2024})},\ \Eprint {https://arxiv.org/abs/2307.03236} {arXiv:2307.03236 [quant-ph]} \BibitemShut {NoStop}%
\bibitem [{\citenamefont {Bauer}\ \emph {et~al.}(2023)\citenamefont {Bauer} \emph {et~al.}}]{Bauer:2022hpo}%
  \BibitemOpen
  \bibfield  {author} {\bibinfo {author} {\bibfnamefont {C.~W.}\ \bibnamefont {Bauer}} \emph {et~al.},\ }\bibfield  {title} {\bibinfo {title} {{Quantum Simulation for High-Energy Physics}},\ }\href {https://doi.org/10.1103/PRXQuantum.4.027001} {\bibfield  {journal} {\bibinfo  {journal} {PRX Quantum}\ }\textbf {\bibinfo {volume} {4}},\ \bibinfo {pages} {027001} (\bibinfo {year} {2023})},\ \Eprint {https://arxiv.org/abs/2204.03381} {arXiv:2204.03381 [quant-ph]} \BibitemShut {NoStop}%
\bibitem [{\citenamefont {Davoudi}\ \emph {et~al.}(2022)\citenamefont {Davoudi} \emph {et~al.}}]{Davoudi:2022bnl}%
  \BibitemOpen
  \bibfield  {author} {\bibinfo {author} {\bibfnamefont {Z.}~\bibnamefont {Davoudi}} \emph {et~al.},\ }\bibfield  {title} {\bibinfo {title} {{Report of the Snowmass 2021 Topical Group on Lattice Gauge Theory}},\ }in\ \href@noop {} {\emph {\bibinfo {booktitle} {{Snowmass 2021}}}}\ (\bibinfo {year} {2022})\ \Eprint {https://arxiv.org/abs/2209.10758} {arXiv:2209.10758 [hep-lat]} \BibitemShut {NoStop}%
\bibitem [{\citenamefont {Catterall}\ \emph {et~al.}(2022)\citenamefont {Catterall} \emph {et~al.}}]{Catterall:2022wjq}%
  \BibitemOpen
  \bibfield  {author} {\bibinfo {author} {\bibfnamefont {S.}~\bibnamefont {Catterall}} \emph {et~al.},\ }\bibfield  {title} {\bibinfo {title} {{Report of the Snowmass 2021 Theory Frontier Topical Group on Quantum Information Science}},\ }in\ \href {https://doi.org/10.2172/1892238} {\emph {\bibinfo {booktitle} {{Snowmass 2021}}}}\ (\bibinfo {year} {2022})\ \Eprint {https://arxiv.org/abs/2209.14839} {arXiv:2209.14839 [quant-ph]} \BibitemShut {NoStop}%
\bibitem [{\citenamefont {Jordan}\ \emph {et~al.}(2012)\citenamefont {Jordan}, \citenamefont {Lee},\ and\ \citenamefont {Preskill}}]{Jordan:2012xnu}%
  \BibitemOpen
  \bibfield  {author} {\bibinfo {author} {\bibfnamefont {S.~P.}\ \bibnamefont {Jordan}}, \bibinfo {author} {\bibfnamefont {K.~S.~M.}\ \bibnamefont {Lee}},\ and\ \bibinfo {author} {\bibfnamefont {J.}~\bibnamefont {Preskill}},\ }\bibfield  {title} {\bibinfo {title} {{Quantum Algorithms for Quantum Field Theories}},\ }\href {https://doi.org/10.1126/science.1217069} {\bibfield  {journal} {\bibinfo  {journal} {Science}\ }\textbf {\bibinfo {volume} {336}},\ \bibinfo {pages} {1130} (\bibinfo {year} {2012})},\ \Eprint {https://arxiv.org/abs/1111.3633} {arXiv:1111.3633 [quant-ph]} \BibitemShut {NoStop}%
\bibitem [{\citenamefont {Jordan}\ \emph {et~al.}(2014{\natexlab{a}})\citenamefont {Jordan}, \citenamefont {Lee},\ and\ \citenamefont {Preskill}}]{Jordan:2011ci}%
  \BibitemOpen
  \bibfield  {author} {\bibinfo {author} {\bibfnamefont {S.~P.}\ \bibnamefont {Jordan}}, \bibinfo {author} {\bibfnamefont {K.~S.~M.}\ \bibnamefont {Lee}},\ and\ \bibinfo {author} {\bibfnamefont {J.}~\bibnamefont {Preskill}},\ }\bibfield  {title} {\bibinfo {title} {{Quantum Computation of Scattering in Scalar Quantum Field Theories}},\ }\href@noop {} {\bibfield  {journal} {\bibinfo  {journal} {Quant. Inf. Comput.}\ }\textbf {\bibinfo {volume} {14}},\ \bibinfo {pages} {1014} (\bibinfo {year} {2014}{\natexlab{a}})},\ \Eprint {https://arxiv.org/abs/1112.4833} {arXiv:1112.4833 [hep-th]} \BibitemShut {NoStop}%
\bibitem [{\citenamefont {Jordan}\ \emph {et~al.}(2014{\natexlab{b}})\citenamefont {Jordan}, \citenamefont {Lee},\ and\ \citenamefont {Preskill}}]{Jordan:2014tma}%
  \BibitemOpen
  \bibfield  {author} {\bibinfo {author} {\bibfnamefont {S.~P.}\ \bibnamefont {Jordan}}, \bibinfo {author} {\bibfnamefont {K.~S.~M.}\ \bibnamefont {Lee}},\ and\ \bibinfo {author} {\bibfnamefont {J.}~\bibnamefont {Preskill}},\ }\bibfield  {title} {\bibinfo {title} {{Quantum Algorithms for Fermionic Quantum Field Theories}},\ }\href@noop {} {\  (\bibinfo {year} {2014}{\natexlab{b}})},\ \Eprint {https://arxiv.org/abs/1404.7115} {arXiv:1404.7115 [hep-th]} \BibitemShut {NoStop}%
\bibitem [{\citenamefont {Jha}\ \emph {et~al.}(2024)\citenamefont {Jha}, \citenamefont {Milsted}, \citenamefont {Neuenfeld}, \citenamefont {Preskill},\ and\ \citenamefont {Vieira}}]{Jha:2024jan}%
  \BibitemOpen
  \bibfield  {author} {\bibinfo {author} {\bibfnamefont {R.~G.}\ \bibnamefont {Jha}}, \bibinfo {author} {\bibfnamefont {A.}~\bibnamefont {Milsted}}, \bibinfo {author} {\bibfnamefont {D.}~\bibnamefont {Neuenfeld}}, \bibinfo {author} {\bibfnamefont {J.}~\bibnamefont {Preskill}},\ and\ \bibinfo {author} {\bibfnamefont {P.}~\bibnamefont {Vieira}},\ }\bibfield  {title} {\bibinfo {title} {{Real-Time Scattering in Ising Field Theory using Matrix Product States}},\ }\href@noop {} {\  (\bibinfo {year} {2024})},\ \Eprint {https://arxiv.org/abs/2411.13645} {arXiv:2411.13645 [hep-th]} \BibitemShut {NoStop}%
\bibitem [{\citenamefont {Gustafson}\ \emph {et~al.}(2019)\citenamefont {Gustafson}, \citenamefont {Meurice},\ and\ \citenamefont {Unmuth-Yockey}}]{Gustafson:2019mpk}%
  \BibitemOpen
  \bibfield  {author} {\bibinfo {author} {\bibfnamefont {E.}~\bibnamefont {Gustafson}}, \bibinfo {author} {\bibfnamefont {Y.}~\bibnamefont {Meurice}},\ and\ \bibinfo {author} {\bibfnamefont {J.}~\bibnamefont {Unmuth-Yockey}},\ }\bibfield  {title} {\bibinfo {title} {{Quantum simulation of scattering in the quantum Ising model}},\ }\href {https://doi.org/10.1103/PhysRevD.99.094503} {\bibfield  {journal} {\bibinfo  {journal} {Phys. Rev. D}\ }\textbf {\bibinfo {volume} {99}},\ \bibinfo {pages} {094503} (\bibinfo {year} {2019})},\ \Eprint {https://arxiv.org/abs/1901.05944} {arXiv:1901.05944 [hep-lat]} \BibitemShut {NoStop}%
\bibitem [{\citenamefont {Gustafson}\ \emph {et~al.}(2021)\citenamefont {Gustafson}, \citenamefont {Zhu}, \citenamefont {Dreher}, \citenamefont {Linke},\ and\ \citenamefont {Meurice}}]{Gustafson:2021imb}%
  \BibitemOpen
  \bibfield  {author} {\bibinfo {author} {\bibfnamefont {E.}~\bibnamefont {Gustafson}}, \bibinfo {author} {\bibfnamefont {Y.}~\bibnamefont {Zhu}}, \bibinfo {author} {\bibfnamefont {P.}~\bibnamefont {Dreher}}, \bibinfo {author} {\bibfnamefont {N.~M.}\ \bibnamefont {Linke}},\ and\ \bibinfo {author} {\bibfnamefont {Y.}~\bibnamefont {Meurice}},\ }\bibfield  {title} {\bibinfo {title} {{Real-time quantum calculations of phase shifts using wave packet time delays}},\ }\href {https://doi.org/10.1103/PhysRevD.104.054507} {\bibfield  {journal} {\bibinfo  {journal} {Phys. Rev. D}\ }\textbf {\bibinfo {volume} {104}},\ \bibinfo {pages} {054507} (\bibinfo {year} {2021})},\ \Eprint {https://arxiv.org/abs/2103.06848} {arXiv:2103.06848 [hep-lat]} \BibitemShut {NoStop}%
\bibitem [{\citenamefont {Parks}\ \emph {et~al.}(2024)\citenamefont {Parks}, \citenamefont {Carignan-Dugas}, \citenamefont {Gustafson}, \citenamefont {Meurice},\ and\ \citenamefont {Dreher}}]{Parks:2022kdb}%
  \BibitemOpen
  \bibfield  {author} {\bibinfo {author} {\bibfnamefont {Z.}~\bibnamefont {Parks}}, \bibinfo {author} {\bibfnamefont {A.}~\bibnamefont {Carignan-Dugas}}, \bibinfo {author} {\bibfnamefont {E.}~\bibnamefont {Gustafson}}, \bibinfo {author} {\bibfnamefont {Y.}~\bibnamefont {Meurice}},\ and\ \bibinfo {author} {\bibfnamefont {P.}~\bibnamefont {Dreher}},\ }\bibfield  {title} {\bibinfo {title} {{Applying the noiseless extrapolation error mitigation protocol to calculate real-time quantum field theory scattering phase shifts}},\ }\href {https://doi.org/10.1103/PhysRevD.109.014505} {\bibfield  {journal} {\bibinfo  {journal} {Phys. Rev. D}\ }\textbf {\bibinfo {volume} {109}},\ \bibinfo {pages} {014505} (\bibinfo {year} {2024})},\ \Eprint {https://arxiv.org/abs/2212.05333} {arXiv:2212.05333 [quant-ph]} \BibitemShut {NoStop}%
\bibitem [{\citenamefont {Ciavarella}(2020)}]{Ciavarella:2020vqm}%
  \BibitemOpen
  \bibfield  {author} {\bibinfo {author} {\bibfnamefont {A.}~\bibnamefont {Ciavarella}},\ }\bibfield  {title} {\bibinfo {title} {{Algorithm for quantum computation of particle decays}},\ }\href {https://doi.org/10.1103/PhysRevD.102.094505} {\bibfield  {journal} {\bibinfo  {journal} {Phys. Rev. D}\ }\textbf {\bibinfo {volume} {102}},\ \bibinfo {pages} {094505} (\bibinfo {year} {2020})},\ \Eprint {https://arxiv.org/abs/2007.04447} {arXiv:2007.04447 [hep-th]} \BibitemShut {NoStop}%
\bibitem [{\citenamefont {Brice\~no}\ \emph {et~al.}(2024)\citenamefont {Brice\~no}, \citenamefont {Edwards}, \citenamefont {Eaton}, \citenamefont {Gonz\'alez-Arciniegas}, \citenamefont {Pfister},\ and\ \citenamefont {Siopsis}}]{Briceno:2023xcm}%
  \BibitemOpen
  \bibfield  {author} {\bibinfo {author} {\bibfnamefont {R.~A.}\ \bibnamefont {Brice\~no}}, \bibinfo {author} {\bibfnamefont {R.~G.}\ \bibnamefont {Edwards}}, \bibinfo {author} {\bibfnamefont {M.}~\bibnamefont {Eaton}}, \bibinfo {author} {\bibfnamefont {C.}~\bibnamefont {Gonz\'alez-Arciniegas}}, \bibinfo {author} {\bibfnamefont {O.}~\bibnamefont {Pfister}},\ and\ \bibinfo {author} {\bibfnamefont {G.}~\bibnamefont {Siopsis}},\ }\bibfield  {title} {\bibinfo {title} {{Toward coherent quantum computation of scattering amplitudes with a measurement-based photonic quantum processor}},\ }\href {https://doi.org/10.1103/PhysRevResearch.6.043065} {\bibfield  {journal} {\bibinfo  {journal} {Phys. Rev. Res.}\ }\textbf {\bibinfo {volume} {6}},\ \bibinfo {pages} {043065} (\bibinfo {year} {2024})},\ \Eprint {https://arxiv.org/abs/2312.12613} {arXiv:2312.12613 [quant-ph]} \BibitemShut {NoStop}%
\bibitem [{\citenamefont {Iadecola}\ \emph {et~al.}(2024)\citenamefont {Iadecola}, \citenamefont {Sen},\ and\ \citenamefont {Sivertsen}}]{Iadecola:2023uti}%
  \BibitemOpen
  \bibfield  {author} {\bibinfo {author} {\bibfnamefont {T.}~\bibnamefont {Iadecola}}, \bibinfo {author} {\bibfnamefont {S.}~\bibnamefont {Sen}},\ and\ \bibinfo {author} {\bibfnamefont {L.}~\bibnamefont {Sivertsen}},\ }\bibfield  {title} {\bibinfo {title} {{Floquet Insulators and Lattice Fermions}},\ }\href {https://doi.org/10.1103/PhysRevLett.132.136601} {\bibfield  {journal} {\bibinfo  {journal} {Phys. Rev. Lett.}\ }\textbf {\bibinfo {volume} {132}},\ \bibinfo {pages} {136601} (\bibinfo {year} {2024})},\ \Eprint {https://arxiv.org/abs/2306.16463} {arXiv:2306.16463 [quant-ph]} \BibitemShut {NoStop}%
\bibitem [{\citenamefont {Chai}\ \emph {et~al.}(2025{\natexlab{b}})\citenamefont {Chai}, \citenamefont {Crippa}, \citenamefont {Jansen}, \citenamefont {K\"uhn}, \citenamefont {Pascuzzi}, \citenamefont {Tacchino},\ and\ \citenamefont {Tavernelli}}]{Chai:2023qpq}%
  \BibitemOpen
  \bibfield  {author} {\bibinfo {author} {\bibfnamefont {Y.}~\bibnamefont {Chai}}, \bibinfo {author} {\bibfnamefont {A.}~\bibnamefont {Crippa}}, \bibinfo {author} {\bibfnamefont {K.}~\bibnamefont {Jansen}}, \bibinfo {author} {\bibfnamefont {S.}~\bibnamefont {K\"uhn}}, \bibinfo {author} {\bibfnamefont {V.~R.}\ \bibnamefont {Pascuzzi}}, \bibinfo {author} {\bibfnamefont {F.}~\bibnamefont {Tacchino}},\ and\ \bibinfo {author} {\bibfnamefont {I.}~\bibnamefont {Tavernelli}},\ }\bibfield  {title} {\bibinfo {title} {{Fermionic wave packet scattering: a quantum computing approach}},\ }\href {https://doi.org/10.22331/q-2025-02-19-1638} {\bibfield  {journal} {\bibinfo  {journal} {Quantum}\ }\textbf {\bibinfo {volume} {9}},\ \bibinfo {pages} {1638} (\bibinfo {year} {2025}{\natexlab{b}})},\ \Eprint {https://arxiv.org/abs/2312.02272} {arXiv:2312.02272 [quant-ph]} \BibitemShut {NoStop}%
\bibitem [{\citenamefont {Farrell}\ \emph {et~al.}(2024{\natexlab{a}})\citenamefont {Farrell}, \citenamefont {Illa}, \citenamefont {Ciavarella},\ and\ \citenamefont {Savage}}]{Farrell:2023fgd}%
  \BibitemOpen
  \bibfield  {author} {\bibinfo {author} {\bibfnamefont {R.~C.}\ \bibnamefont {Farrell}}, \bibinfo {author} {\bibfnamefont {M.}~\bibnamefont {Illa}}, \bibinfo {author} {\bibfnamefont {A.~N.}\ \bibnamefont {Ciavarella}},\ and\ \bibinfo {author} {\bibfnamefont {M.~J.}\ \bibnamefont {Savage}},\ }\bibfield  {title} {\bibinfo {title} {{Scalable Circuits for Preparing Ground States on Digital Quantum Computers: The Schwinger Model Vacuum on 100 Qubits}},\ }\href {https://doi.org/10.1103/PRXQuantum.5.020315} {\bibfield  {journal} {\bibinfo  {journal} {PRX Quantum}\ }\textbf {\bibinfo {volume} {5}},\ \bibinfo {pages} {020315} (\bibinfo {year} {2024}{\natexlab{a}})},\ \Eprint {https://arxiv.org/abs/2308.04481} {arXiv:2308.04481 [quant-ph]} \BibitemShut {NoStop}%
\bibitem [{\citenamefont {Farrell}\ \emph {et~al.}(2024{\natexlab{b}})\citenamefont {Farrell}, \citenamefont {Illa}, \citenamefont {Ciavarella},\ and\ \citenamefont {Savage}}]{Farrell:2024fit}%
  \BibitemOpen
  \bibfield  {author} {\bibinfo {author} {\bibfnamefont {R.~C.}\ \bibnamefont {Farrell}}, \bibinfo {author} {\bibfnamefont {M.}~\bibnamefont {Illa}}, \bibinfo {author} {\bibfnamefont {A.~N.}\ \bibnamefont {Ciavarella}},\ and\ \bibinfo {author} {\bibfnamefont {M.~J.}\ \bibnamefont {Savage}},\ }\bibfield  {title} {\bibinfo {title} {{Quantum simulations of hadron dynamics in the Schwinger model using 112 qubits}},\ }\href {https://doi.org/10.1103/PhysRevD.109.114510} {\bibfield  {journal} {\bibinfo  {journal} {Phys. Rev. D}\ }\textbf {\bibinfo {volume} {109}},\ \bibinfo {pages} {114510} (\bibinfo {year} {2024}{\natexlab{b}})},\ \Eprint {https://arxiv.org/abs/2401.08044} {arXiv:2401.08044 [quant-ph]} \BibitemShut {NoStop}%
\bibitem [{\citenamefont {Ciavarella}(2025)}]{Ciavarella:2024lsp}%
  \BibitemOpen
  \bibfield  {author} {\bibinfo {author} {\bibfnamefont {A.~N.}\ \bibnamefont {Ciavarella}},\ }\bibfield  {title} {\bibinfo {title} {{String breaking in the heavy quark limit with scalable circuits}},\ }\href {https://doi.org/10.1103/PhysRevD.111.054501} {\bibfield  {journal} {\bibinfo  {journal} {Phys. Rev. D}\ }\textbf {\bibinfo {volume} {111}},\ \bibinfo {pages} {054501} (\bibinfo {year} {2025})},\ \Eprint {https://arxiv.org/abs/2411.05915} {arXiv:2411.05915 [quant-ph]} \BibitemShut {NoStop}%
\bibitem [{\citenamefont {Farrell}\ \emph {et~al.}(2025{\natexlab{b}})\citenamefont {Farrell}, \citenamefont {Illa},\ and\ \citenamefont {Savage}}]{Farrell:2024mgu}%
  \BibitemOpen
  \bibfield  {author} {\bibinfo {author} {\bibfnamefont {R.~C.}\ \bibnamefont {Farrell}}, \bibinfo {author} {\bibfnamefont {M.}~\bibnamefont {Illa}},\ and\ \bibinfo {author} {\bibfnamefont {M.~J.}\ \bibnamefont {Savage}},\ }\bibfield  {title} {\bibinfo {title} {{Steps toward quantum simulations of hadronization and energy loss in dense matter}},\ }\href {https://doi.org/10.1103/PhysRevC.111.015202} {\bibfield  {journal} {\bibinfo  {journal} {Phys. Rev. C}\ }\textbf {\bibinfo {volume} {111}},\ \bibinfo {pages} {015202} (\bibinfo {year} {2025}{\natexlab{b}})},\ \Eprint {https://arxiv.org/abs/2405.06620} {arXiv:2405.06620 [quant-ph]} \BibitemShut {NoStop}%
\bibitem [{\citenamefont {Gustafson}\ \emph {et~al.}(2024)\citenamefont {Gustafson} \emph {et~al.}}]{Gustafson:2024bww}%
  \BibitemOpen
  \bibfield  {author} {\bibinfo {author} {\bibfnamefont {E.}~\bibnamefont {Gustafson}} \emph {et~al.},\ }\bibfield  {title} {\bibinfo {title} {{Surrogate Constructed Scalable Circuits ADAPT-VQE in the Schwinger model}},\ }\href@noop {} {\  (\bibinfo {year} {2024})},\ \Eprint {https://arxiv.org/abs/2408.12641} {arXiv:2408.12641 [quant-ph]} \BibitemShut {NoStop}%
\bibitem [{\citenamefont {Davoudi}\ \emph {et~al.}(2024)\citenamefont {Davoudi}, \citenamefont {Hsieh},\ and\ \citenamefont {Kadam}}]{Davoudi:2024wyv}%
  \BibitemOpen
  \bibfield  {author} {\bibinfo {author} {\bibfnamefont {Z.}~\bibnamefont {Davoudi}}, \bibinfo {author} {\bibfnamefont {C.-C.}\ \bibnamefont {Hsieh}},\ and\ \bibinfo {author} {\bibfnamefont {S.~V.}\ \bibnamefont {Kadam}},\ }\bibfield  {title} {\bibinfo {title} {{Scattering wave packets of hadrons in gauge theories: Preparation on a quantum computer}},\ }\href {https://doi.org/10.22331/q-2024-11-11-1520} {\bibfield  {journal} {\bibinfo  {journal} {Quantum}\ }\textbf {\bibinfo {volume} {8}},\ \bibinfo {pages} {1520} (\bibinfo {year} {2024})},\ \Eprint {https://arxiv.org/abs/2402.00840} {arXiv:2402.00840 [quant-ph]} \BibitemShut {NoStop}%
\bibitem [{\citenamefont {Liu}\ \emph {et~al.}(2022)\citenamefont {Liu}, \citenamefont {Li}, \citenamefont {Zheng}, \citenamefont {Yuan},\ and\ \citenamefont {Sun}}]{Liu:2021otn}%
  \BibitemOpen
  \bibfield  {author} {\bibinfo {author} {\bibfnamefont {J.}~\bibnamefont {Liu}}, \bibinfo {author} {\bibfnamefont {Z.}~\bibnamefont {Li}}, \bibinfo {author} {\bibfnamefont {H.}~\bibnamefont {Zheng}}, \bibinfo {author} {\bibfnamefont {X.}~\bibnamefont {Yuan}},\ and\ \bibinfo {author} {\bibfnamefont {J.}~\bibnamefont {Sun}},\ }\bibfield  {title} {\bibinfo {title} {{Towards a variational Jordan\textendash{}Lee\textendash{}Preskill quantum algorithm}},\ }\href {https://doi.org/10.1088/2632-2153/aca06b} {\bibfield  {journal} {\bibinfo  {journal} {Mach. Learn. Sci. Tech.}\ }\textbf {\bibinfo {volume} {3}},\ \bibinfo {pages} {045030} (\bibinfo {year} {2022})},\ \Eprint {https://arxiv.org/abs/2109.05547} {arXiv:2109.05547 [quant-ph]} \BibitemShut {NoStop}%
\bibitem [{\citenamefont {Zemlevskiy}(2024)}]{Zemlevskiy:2024vxt}%
  \BibitemOpen
  \bibfield  {author} {\bibinfo {author} {\bibfnamefont {N.~A.}\ \bibnamefont {Zemlevskiy}},\ }\bibfield  {title} {\bibinfo {title} {{Scalable Quantum Simulations of Scattering in Scalar Field Theory on 120 Qubits}},\ }\href@noop {} {\  (\bibinfo {year} {2024})},\ \Eprint {https://arxiv.org/abs/2411.02486} {arXiv:2411.02486 [quant-ph]} \BibitemShut {NoStop}%
\bibitem [{\citenamefont {Pedernales}\ \emph {et~al.}(2014)\citenamefont {Pedernales}, \citenamefont {Di~Candia}, \citenamefont {Egusquiza}, \citenamefont {Casanova},\ and\ \citenamefont {Solano}}]{PhysRevLett.113.020505}%
  \BibitemOpen
  \bibfield  {author} {\bibinfo {author} {\bibfnamefont {J.~S.}\ \bibnamefont {Pedernales}}, \bibinfo {author} {\bibfnamefont {R.}~\bibnamefont {Di~Candia}}, \bibinfo {author} {\bibfnamefont {I.~L.}\ \bibnamefont {Egusquiza}}, \bibinfo {author} {\bibfnamefont {J.}~\bibnamefont {Casanova}},\ and\ \bibinfo {author} {\bibfnamefont {E.}~\bibnamefont {Solano}},\ }\bibfield  {title} {\bibinfo {title} {Efficient quantum algorithm for computing $n$-time correlation functions},\ }\href {https://doi.org/10.1103/PhysRevLett.113.020505} {\bibfield  {journal} {\bibinfo  {journal} {Phys. Rev. Lett.}\ }\textbf {\bibinfo {volume} {113}},\ \bibinfo {pages} {020505} (\bibinfo {year} {2014})}\BibitemShut {NoStop}%
\bibitem [{\citenamefont {Wang}\ \emph {et~al.}(2025)\citenamefont {Wang}, \citenamefont {Xiong}, \citenamefont {Cai},\ and\ \citenamefont {Yuan}}]{wang2025computingntimecorrelationfunctions}%
  \BibitemOpen
  \bibfield  {author} {\bibinfo {author} {\bibfnamefont {X.}~\bibnamefont {Wang}}, \bibinfo {author} {\bibfnamefont {L.}~\bibnamefont {Xiong}}, \bibinfo {author} {\bibfnamefont {X.}~\bibnamefont {Cai}},\ and\ \bibinfo {author} {\bibfnamefont {X.}~\bibnamefont {Yuan}},\ }\href {https://arxiv.org/abs/2504.12975} {\bibinfo {title} {Computing $n$-time correlation functions without ancilla qubits}} (\bibinfo {year} {2025}),\ \Eprint {https://arxiv.org/abs/2504.12975} {arXiv:2504.12975 [quant-ph]} \BibitemShut {NoStop}%
\bibitem [{\citenamefont {Farrell}\ \emph {et~al.}(2023)\citenamefont {Farrell}, \citenamefont {Chernyshev}, \citenamefont {Powell}, \citenamefont {Zemlevskiy}, \citenamefont {Illa},\ and\ \citenamefont {Savage}}]{Farrell:2022vyh}%
  \BibitemOpen
  \bibfield  {author} {\bibinfo {author} {\bibfnamefont {R.~C.}\ \bibnamefont {Farrell}}, \bibinfo {author} {\bibfnamefont {I.~A.}\ \bibnamefont {Chernyshev}}, \bibinfo {author} {\bibfnamefont {S.~J.~M.}\ \bibnamefont {Powell}}, \bibinfo {author} {\bibfnamefont {N.~A.}\ \bibnamefont {Zemlevskiy}}, \bibinfo {author} {\bibfnamefont {M.}~\bibnamefont {Illa}},\ and\ \bibinfo {author} {\bibfnamefont {M.~J.}\ \bibnamefont {Savage}},\ }\bibfield  {title} {\bibinfo {title} {{Preparations for quantum simulations of quantum chromodynamics in 1+1 dimensions. II. Single-baryon \ensuremath{\beta}-decay in real time}},\ }\href {https://doi.org/10.1103/PhysRevD.107.054513} {\bibfield  {journal} {\bibinfo  {journal} {Phys. Rev. D}\ }\textbf {\bibinfo {volume} {107}},\ \bibinfo {pages} {054513} (\bibinfo {year} {2023})},\ \Eprint {https://arxiv.org/abs/2209.10781} {arXiv:2209.10781 [quant-ph]} \BibitemShut {NoStop}%
\bibitem [{Note2()}]{Note2}%
  \BibitemOpen
  \bibinfo {note} {Given our interest in finite-volume artifacts, we will not discuss ultraviolet (UV) divergences in this work. Since both finite-volume and infinite-volume diagrams share the same UV structure, finite-volume corrections are UV finite.}\BibitemShut {Stop}%
\bibitem [{\citenamefont {Coleman}(2018)}]{Coleman:2018mew}%
  \BibitemOpen
  \bibfield  {author} {\bibinfo {author} {\bibfnamefont {S.}~\bibnamefont {Coleman}},\ }\href {https://doi.org/10.1142/9371} {\emph {\bibinfo {title} {{Lectures of Sidney Coleman on Quantum Field Theory}}}},\ edited by\ \bibinfo {editor} {\bibfnamefont {B.~G.-g.}\ \bibnamefont {Chen}}, \bibinfo {editor} {\bibfnamefont {D.}~\bibnamefont {Derbes}}, \bibinfo {editor} {\bibfnamefont {D.}~\bibnamefont {Griffiths}}, \bibinfo {editor} {\bibfnamefont {B.}~\bibnamefont {Hill}}, \bibinfo {editor} {\bibfnamefont {R.}~\bibnamefont {Sohn}},\ and\ \bibinfo {editor} {\bibfnamefont {Y.-S.}\ \bibnamefont {Ting}}\ (\bibinfo  {publisher} {WSP},\ \bibinfo {address} {Hackensack},\ \bibinfo {year} {2018})\BibitemShut {NoStop}%
\bibitem [{\citenamefont {Weinzierl}(2022)}]{Weinzierl:2022eaz}%
  \BibitemOpen
  \bibfield  {author} {\bibinfo {author} {\bibfnamefont {S.}~\bibnamefont {Weinzierl}},\ }\href {https://doi.org/10.1007/978-3-030-99558-4} {\emph {\bibinfo {title} {{Feynman Integrals. A Comprehensive Treatment for Students and Researchers}}}},\ UNITEXT for Physics\ (\bibinfo  {publisher} {Springer},\ \bibinfo {year} {2022})\ \Eprint {https://arxiv.org/abs/2201.03593} {arXiv:2201.03593 [hep-th]} \BibitemShut {NoStop}%
\bibitem [{Note3()}]{Note3}%
  \BibitemOpen
  \bibinfo {note} {The reader might find it useful to note that $g\Delta $ is the second Symanzik polynomial.}\BibitemShut {Stop}%
\bibitem [{\citenamefont {Davoudi}\ and\ \citenamefont {Kadam}(2020)}]{Davoudi:2020xdv}%
  \BibitemOpen
  \bibfield  {author} {\bibinfo {author} {\bibfnamefont {Z.}~\bibnamefont {Davoudi}}\ and\ \bibinfo {author} {\bibfnamefont {S.~V.}\ \bibnamefont {Kadam}},\ }\bibfield  {title} {\bibinfo {title} {{Two-neutrino double-$\beta$ decay in pionless effective field theory from a Euclidean finite-volume correlation function}},\ }\href {https://doi.org/10.1103/PhysRevD.102.114521} {\bibfield  {journal} {\bibinfo  {journal} {Phys. Rev. D}\ }\textbf {\bibinfo {volume} {102}},\ \bibinfo {pages} {114521} (\bibinfo {year} {2020})},\ \Eprint {https://arxiv.org/abs/2007.15542} {arXiv:2007.15542 [hep-lat]} \BibitemShut {NoStop}%
\bibitem [{\citenamefont {Baroni}\ \emph {et~al.}(2019)\citenamefont {Baroni}, \citenamefont {Brice\~no}, \citenamefont {Hansen},\ and\ \citenamefont {Ortega-Gama}}]{Baroni:2018iau}%
  \BibitemOpen
  \bibfield  {author} {\bibinfo {author} {\bibfnamefont {A.}~\bibnamefont {Baroni}}, \bibinfo {author} {\bibfnamefont {R.~A.}\ \bibnamefont {Brice\~no}}, \bibinfo {author} {\bibfnamefont {M.~T.}\ \bibnamefont {Hansen}},\ and\ \bibinfo {author} {\bibfnamefont {F.~G.}\ \bibnamefont {Ortega-Gama}},\ }\bibfield  {title} {\bibinfo {title} {{Form factors of two-hadron states from a covariant finite-volume formalism}},\ }\href {https://doi.org/10.1103/PhysRevD.100.034511} {\bibfield  {journal} {\bibinfo  {journal} {Phys. Rev. D}\ }\textbf {\bibinfo {volume} {100}},\ \bibinfo {pages} {034511} (\bibinfo {year} {2019})},\ \Eprint {https://arxiv.org/abs/1812.10504} {arXiv:1812.10504 [hep-lat]} \BibitemShut {NoStop}%
\bibitem [{\citenamefont {Brice\~no}\ and\ \citenamefont {Hansen}(2016)}]{Briceno:2015tza}%
  \BibitemOpen
  \bibfield  {author} {\bibinfo {author} {\bibfnamefont {R.~A.}\ \bibnamefont {Brice\~no}}\ and\ \bibinfo {author} {\bibfnamefont {M.~T.}\ \bibnamefont {Hansen}},\ }\bibfield  {title} {\bibinfo {title} {{Relativistic, model-independent, multichannel $2\to 2$ transition amplitudes in a finite volume}},\ }\href {https://doi.org/10.1103/PhysRevD.94.013008} {\bibfield  {journal} {\bibinfo  {journal} {Phys. Rev. D}\ }\textbf {\bibinfo {volume} {94}},\ \bibinfo {pages} {013008} (\bibinfo {year} {2016})},\ \Eprint {https://arxiv.org/abs/1509.08507} {arXiv:1509.08507 [hep-lat]} \BibitemShut {NoStop}%
\bibitem [{\citenamefont {Briceno}\ and\ \citenamefont {Davoudi}(2013)}]{Briceno:2012yi}%
  \BibitemOpen
  \bibfield  {author} {\bibinfo {author} {\bibfnamefont {R.~A.}\ \bibnamefont {Briceno}}\ and\ \bibinfo {author} {\bibfnamefont {Z.}~\bibnamefont {Davoudi}},\ }\bibfield  {title} {\bibinfo {title} {{Moving multichannel systems in a finite volume with application to proton-proton fusion}},\ }\href {https://doi.org/10.1103/PhysRevD.88.094507} {\bibfield  {journal} {\bibinfo  {journal} {Phys. Rev. D}\ }\textbf {\bibinfo {volume} {88}},\ \bibinfo {pages} {094507} (\bibinfo {year} {2013})},\ \Eprint {https://arxiv.org/abs/1204.1110} {arXiv:1204.1110 [hep-lat]} \BibitemShut {NoStop}%
\bibitem [{\citenamefont {Lozano}\ \emph {et~al.}(2022)\citenamefont {Lozano}, \citenamefont {Mei\ss{}ner}, \citenamefont {Romero-L\'opez}, \citenamefont {Rusetsky},\ and\ \citenamefont {Schierholz}}]{Lozano:2022kfz}%
  \BibitemOpen
  \bibfield  {author} {\bibinfo {author} {\bibfnamefont {J.}~\bibnamefont {Lozano}}, \bibinfo {author} {\bibfnamefont {U.-G.}\ \bibnamefont {Mei\ss{}ner}}, \bibinfo {author} {\bibfnamefont {F.}~\bibnamefont {Romero-L\'opez}}, \bibinfo {author} {\bibfnamefont {A.}~\bibnamefont {Rusetsky}},\ and\ \bibinfo {author} {\bibfnamefont {G.}~\bibnamefont {Schierholz}},\ }\bibfield  {title} {\bibinfo {title} {{Resonance form factors from finite-volume correlation functions with the external field method}},\ }\href {https://doi.org/10.1007/JHEP10(2022)106} {\bibfield  {journal} {\bibinfo  {journal} {JHEP}\ }\textbf {\bibinfo {volume} {10}},\ \bibinfo {pages} {106}},\ \Eprint {https://arxiv.org/abs/2205.11316} {arXiv:2205.11316 [hep-lat]} \BibitemShut {NoStop}%
\bibitem [{\citenamefont {Mizera}\ and\ \citenamefont {Telen}(2022)}]{Mizera:2021icv}%
  \BibitemOpen
  \bibfield  {author} {\bibinfo {author} {\bibfnamefont {S.}~\bibnamefont {Mizera}}\ and\ \bibinfo {author} {\bibfnamefont {S.}~\bibnamefont {Telen}},\ }\bibfield  {title} {\bibinfo {title} {{Landau discriminants}},\ }\href {https://doi.org/10.1007/JHEP08(2022)200} {\bibfield  {journal} {\bibinfo  {journal} {JHEP}\ }\textbf {\bibinfo {volume} {08}},\ \bibinfo {pages} {200}},\ \Eprint {https://arxiv.org/abs/2109.08036} {arXiv:2109.08036 [math-ph]} \BibitemShut {NoStop}%
\bibitem [{\citenamefont {Kim}\ \emph {et~al.}(2005)\citenamefont {Kim}, \citenamefont {Sachrajda},\ and\ \citenamefont {Sharpe}}]{Kim:2005gf}%
  \BibitemOpen
  \bibfield  {author} {\bibinfo {author} {\bibfnamefont {C.~h.}\ \bibnamefont {Kim}}, \bibinfo {author} {\bibfnamefont {C.~T.}\ \bibnamefont {Sachrajda}},\ and\ \bibinfo {author} {\bibfnamefont {S.~R.}\ \bibnamefont {Sharpe}},\ }\bibfield  {title} {\bibinfo {title} {{Finite-volume effects for two-hadron states in moving frames}},\ }\href {https://doi.org/10.1016/j.nuclphysb.2005.08.029} {\bibfield  {journal} {\bibinfo  {journal} {Nucl. Phys. B}\ }\textbf {\bibinfo {volume} {727}},\ \bibinfo {pages} {218} (\bibinfo {year} {2005})},\ \Eprint {https://arxiv.org/abs/hep-lat/0507006} {arXiv:hep-lat/0507006} \BibitemShut {NoStop}%
\bibitem [{\citenamefont {Briceno}\ \emph {et~al.}(2014)\citenamefont {Briceno}, \citenamefont {Davoudi}, \citenamefont {Luu},\ and\ \citenamefont {Savage}}]{Briceno:2013hya}%
  \BibitemOpen
  \bibfield  {author} {\bibinfo {author} {\bibfnamefont {R.~A.}\ \bibnamefont {Briceno}}, \bibinfo {author} {\bibfnamefont {Z.}~\bibnamefont {Davoudi}}, \bibinfo {author} {\bibfnamefont {T.~C.}\ \bibnamefont {Luu}},\ and\ \bibinfo {author} {\bibfnamefont {M.~J.}\ \bibnamefont {Savage}},\ }\bibfield  {title} {\bibinfo {title} {{Two-Baryon Systems with Twisted Boundary Conditions}},\ }\href {https://doi.org/10.1103/PhysRevD.89.074509} {\bibfield  {journal} {\bibinfo  {journal} {Phys. Rev. D}\ }\textbf {\bibinfo {volume} {89}},\ \bibinfo {pages} {074509} (\bibinfo {year} {2014})},\ \Eprint {https://arxiv.org/abs/1311.7686} {arXiv:1311.7686 [hep-lat]} \BibitemShut {NoStop}%
\end{thebibliography}
\end{document}